\documentclass[floatfix, nofootinbib]{emulateapj}
\usepackage{epsfig,subfigure}
\usepackage{color}

\definecolor{purple}{rgb}{1,0,1}

\newcommand{\lcdm}{$\Lambda$CDM}
\newcommand{\hmpc}{$h^{-1}$Mpc}
\newcommand{\hmpccub}{$h^{-3}$Mpc$^3$}
\newcommand{\zobov}{{\tt ZOBOV}}
\newcommand{\healpix}{{\tt HEALPix}}

\bibpunct[; ]{(}{)}{;}{a}{}{,}

\begin{document}

\title{A public void catalog from the 
       SDSS DR7 Galaxy Redshift Surveys based on the watershed transform}
\author{ P.~M.~Sutter$^{1,2,3,4}$, 
         Guilhem Lavaux$^{5,6}$,
         Benjamin~D.~Wandelt$^{2,3,1,7}$, and
         David~H.~Weinberg$^{4,8}$\\
 {~}\\
$^{1}$ Department of Physics, University of Illinois at Urbana-Champaign, Urbana, IL 61801\\
$^{2}$ UPMC Univ Paris 06, UMR7095, Institut d'Astrophysique de Paris, F-75014, Paris, France \\
$^{3}$ CNRS, UMR7095, Institut d'Astrophysique de Paris, F-75014, Paris, France 
\\
$^{4}$ Center for Cosmology and Astro-Particle Physics, Ohio State University, Columbus, OH 43210\\
$^{5}$ Department of Physics \& Astronomy, University of Waterloo, Waterloo,
ON,  N2L 3G1 Canada \\
$^{6}$ Perimeter Institute for Theoretical Physics, 31 Caroline St. N.,
Waterloo, ON, N2L 2Y5, Canada \\
$^{7}$ Department of Astronomy, University of Illinois at Urbana-Champaign, Urbana, IL 61801\\
$^{8}$ Department of Astronomy, Ohio State University, Columbus, OH 43210\\
}
%\thanks{Email: psutter2@illinois.edu}

\begin{abstract}
We produce the most comprehensive public void catalog to date using 
the Sloan Digital Sky Survey Data Release 7 
main sample out to redshift $z=0.2$ and the 
luminous red galaxy sample out to $z=0.44$.
Using a modified version of the parameter-free void finder {\tt ZOBOV}, we 
fully take into account the presence of the survey boundary and masks.
Our strategy for finding voids is thus appropriate for any survey 
configuration.
We produce two distinct catalogs: a complete catalog including voids near 
any masks, which would be
 appropriate for void galaxy surveys, and a bias-free catalog 
of voids away from any masks, which is necessary for analyses that 
require a fair sampling of void shapes and alignments.
Our discovered voids have effective radii from 5 to 135~\hmpc. 
We discuss basic catalog statistics such as number 
counts and redshift distributions and describe some additional 
data products derived from our catalog, such as radial density profiles 
and projected density maps. 
We find that radial profiles of stacked voids show a 
qualitatively similar behavior 
across nearly two decades of void radii and throughout the full 
redshift range.
\end{abstract}


\maketitle

%-------------------------------------------------------------------------------
\section{Introduction}
\label{sec:introduction}

The hierarchical clustering of matter in the universe naturally leads to large
underdense regions, called voids. Indeed, the presence of voids in the
large-scale distribution of galaxies was one of the early predictions of cold
dark matter cosmological theories~\citep{Hausman1983}, and the discovery of
voids in some of the first galaxy redshift 
surveys~\citep{Gregory1978,Kirshner1981} quickly provided
a rich source of interest.  Today, galaxy surveys, such as the Void Galaxy 
Survey~\citep{VandeWeygaertR.2011} and the Las Companas Redshift 
Survey~\citep{Muller2000} 
routinely find and
characterize both voids themselves and their contents for useful astrophysical
and cosmological information (see~\citealt{Thompson2011} 
for a review).  

A combination of observations and simulations gives a coherent picture of void
properties. In the cosmological constant 
plus cold dark matter ($\Lambda$CDM) picture of
cosmic evolution, voids --- usually defined to have densities 
of 10-20\% the cosmic mean --- have 
characteristic radii of 10-40~\hmpc, with the 
smallest identifiable voids in the local universe having radii $\sim7$~\hmpc
~\citep{Tikhonov2006}. 
Early structure formation simulations successfully
reconstructed observed voids~\citep{Hoffman1982, White1987}. Later studies of
voids from surveys such as IRAS~\citep{Plionis2002}, the 2dF Galaxy
Redshift Survey~\citep{Hoyle2004}, and the Sloan Digital Sky
Survey~\citep{Pan2011} confirmed these properties.  
Semi-analytic
models~\citep{Benson2003, Tinker2009} and large-scale \emph{ab initio}
simulations~\citep{Dubinski1993, Colberg2005, Ceccarelli2006, Park2007, Kreckel2011} have
further elucidated the evolution, internal structure, and distribution of
voids.

Since voids are nearly empty, their dynamics are dominated by 
dark energy. Thus, they may provide crucial probes of
primordial density fluctuations~\citep{Sahni1994},
fifth forces~\citep{Li2009},
and $F(r)$ gravity models~\citep{Li2012}.
The Alcock-Paczynski test~\citep{Alcock1979} can be applied to 
measurements of void ellipticities, directly probing the expansion history 
of the universe~\citep{Ryden1995, Ryden1996, Park2007, Biswas2010, 
LavauxGuilhem2011}.
The internal structure of voids behaves as a universe in miniature, allowing 
for probes of the history of dark energy~\citep{Gottlober2003, Goldberg2004}.
The ellipticity distribution 
of voids can provide insights into the growth of structure and 
the correctness of General Relativity~\citep{Shoji,Lavaux2010}.
Void orientation and spin statistics reveal information on large-scale 
tidal fields~\citep{Lee2006, Platen2008}.
Understanding the locations and sizes of voids is also crucial 
for cosmic microwave background (CMB) missions, since they affect the 
CMB signal via the integrated Sachs-Wolfe 
effect~\citep{Thompson1987, Vadas1998, Cruz2008, Gurzadyan2009, Granett2008}.

The reliability of the above conclusions and predictions 
rests on the ability to robustly produce void catalogs from galaxy surveys.
While void finders are well-studied in the context of the large-scale 
dark matter simulations~\citep[e.g.,][]{Colberg2005, Colberg2008}, 
few are applied directly to large-scale redshift surveys.
The largest void catalogs previous to this work use void finders
that rely on
overlapping spheres of underdensities~\citep{Hoyle2004, Pan2011}. 
While simple to apply, this approach 
fails to capture the full geometry of the voids and relies on 
finely-tuned parameters to correctly capture them.
Additionally, previous works ignore the presence of survey 
boundaries and masks and do not extend to the full redshift range 
of the available surveys.

In this work, we describe our techniques for accounting for 
biases due to the presence of a survey boundary and masks.
We employ these techniques along with 
a modified version of the parameter-free void finder 
{\tt ZOBOV}~\citep{Neyrinck2008, LavauxGuilhem2011} to produce a
catalog of voids from both the main sample~\citep{Strauss2002} and
luminous red galaxy (LRG)~\citep{Eisenstein2001} sample of
the Sloan Digital Sky Survey (SDSS) Data Release 7~\citep{Abazajian2009}.
The void catalog we produce will be useful for many void-based 
astrophysical and cosmological studies, as already noted. 
This void catalog extends to higher redshifts than other 
catalogs~\citep[e.g.,][]{Plionis2002,Hoyle2004,Pan2011,Sousbie2011,VandeWeygaertR.2011} 
and is the first to include not only 
main sample galaxies but also LRGs.
While the voids we will identify in the LRG sample are topologically 
consistent (based on the tessellation and watershed procedures in {\tt ZOBOV}), 
they may not fully 
correspond to underdensities in the cosmological sense due to undersampling 
of the density field and galaxy biasing effects. We will return to this 
discussion in Section~\ref{sec:conclusions}.

We begin in Section~\ref{sec:samples} with a presentation of
 our selection of data samples 
from the SDSS catalog. In Section~\ref{sec:finding} we describe our 
modifications to {\tt ZOBOV} to handle the survey boundary and masks as 
well as our process for eliminating alignment biases in the 
void catalog.
We characterize the demographics of our void catalog, including 
redshift-dependent number counts and radial density profiles, in 
Section~\ref{sec:properties}.
We provide two examples of derived data products from the void catalog:
radial profiles of stacked voids in Section~\ref{sec:radial} and
projected density maps of stacked voids in 
Section~\ref{sec:projected}.
We discuss the potential for future applications in 
Section~\ref{sec:conclusions} and provide details of the layout of the 
public void catalog in the Appendix.

%-------------------------------------------------------------------------------
\section{Data samples}
\label{sec:samples}

We identify voids in both the SDSS main galaxy redshift 
survey~\citep{Strauss2002} and the SDSS luminous red galaxy (LRG) 
redshift survey~\citep{Eisenstein2001}.
We take our main galaxy sample from the New York University Value-Added 
Galaxy Catalog~\citep{Blanton2005} which cross-matches galaxies 
from SDSS~\citep{Abazajian2009} with other surveys using 
improved photometric calibrations~\citep{Padmanabhan2008}. 
We draw our samples from the \emph{full1} catalog, which
enforces evolution- and $K$-corrected magnitude limits of $-23 < M_r < -17$.
This catalog extends to $z \sim 0.3$ and 
contains 671,451 galaxies.
We take the LRG catalog of~\citet{Kazin2010}, which 
extends from $z=0.16$ to $z=0.47$ and enforces magnitude limits 
of $-23.2 < M_r < -21.2$. This catalog contains 105,831 galaxies.

The properties of a void necessarily depend on the galaxy distribution 
used to define it.
We wish to have statistically uniform void populations, so for this 
work we choose statistically uniform,
volume-limited subsets of the above catalog.
Although 
our algorithm does not strictly require it, 
using volume-limited samples maintains similar effects of 
shot noise and galaxy bias throughout each void catalog
and in principle allows one to make 
prompt comparison 
to predictions based on halo occupation models fit to galaxy correlation 
functions~\citep{Berlind2002, Zahavi2011}.

\begin{figure*} 
  \centering 
  {\includegraphics[width=0.49\textwidth]{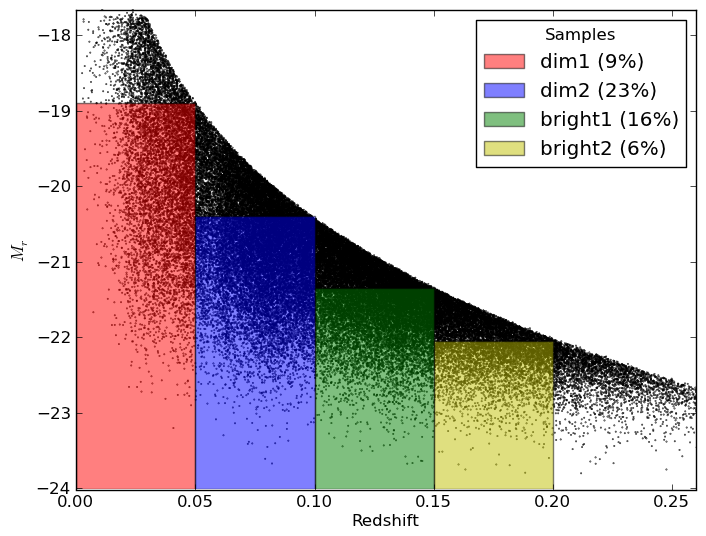}}
  {\includegraphics[width=0.49\textwidth]{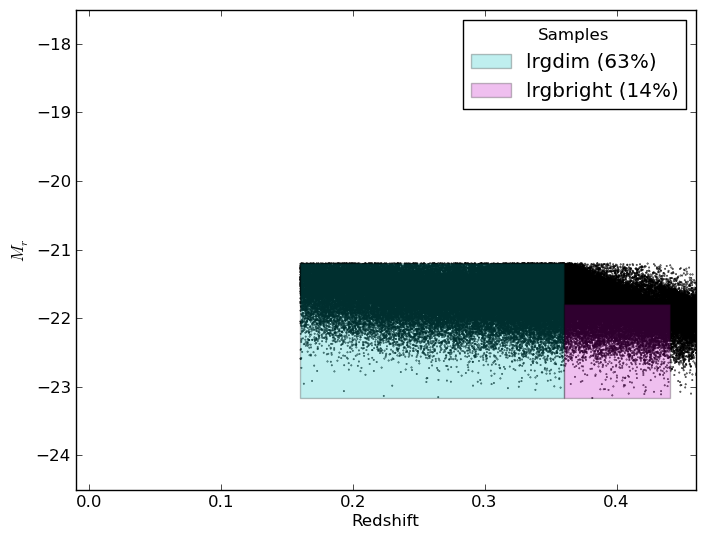}}
  \caption{\emph{Selection of volume-limited samples.} 
           Volume-limited samples of the main sample (left) and LRGs (right). 
           Catalog galaxies are shown in black. We reduce the main sample
           galaxies by a factor of 10 to improve clarity.
           We show the percentage of galaxies in each sample with the 
           sample name.}
\label{fig:samples}
\end{figure*}

For the main sample, we apply evolution and $K$-corrections and 
compute absolute magnitudes $M_r$ assuming
a cosmology consistent with the latest WMAP 7-year 
results~\citep{Komatsu2011}: $\Omega_M=0.27$, $\Omega_\Lambda=0.73$, 
and $h=0.71$.
Figure~\ref{fig:samples} shows the absolute $r$-band magnitude versus redshift
for each catalog and illustrates 
our chosen samples.
We choose four redshift bins. 
Each redshift range is characterized by a typical galaxy luminosity, 
which we differentiate by the labels~\emph{dim} and~\emph{bright}.
Our redshift bins are:
$0.0 < z < 0.05$, which we label \emph{dim1}, 
$0.05 < z < 0.1$, labeled as \emph{dim2}, 
$0.1 < z < 0.15$, called \emph{bright1},
and $0.15 < z <0.2$, which 
we label \emph{bright2}. 
For the LRGs, we use two volume-limited samples 
which we label here as \emph{lrgdim} and 
\emph{lrgbright}. The LRG catalogs are already nearly volume-limited 
by construction, and details of constructing these two subsamples can be 
found in~\citet{Kazin2010}.
Table~\ref{tab:samples} 
lists the sample name, source catalog, limiting absolute magnitude, 
redshift bound, number of galaxies, and average galaxy spacing.
The average galaxy spacing is $(n_g/V)^{1/3}$, where 
$n_g$ is the number of galaxies within each sample and $V$ is the 
sample volume.

To gather as many voids as possible, we take each sample from 
redshift $z=0.0$ to its given redshift cutoff. However, we 
drop from the sample any void whose center crosses the outer 
redshift limit of an interior sample.
This minimizes boundary effects.

\begin{table*}
\centering
\caption{Volume-limited Samples Used in this Work.}
\tabcolsep=0.11cm
\footnotesize
\begin{tabular}{ccccccc}
  Sample Name & Catalog & $M_{r, {\rm max}}$ & $z_{\rm min}$ & 
              $z_{\rm max}$ & Number of Galaxies & Mean Spacing (\hmpc) \\
  \hline
  \hline
  dim1 & NYU VAGC & -18.9 & 0.0 & 0.05 & 63639 &  3 \\ 
dim2 & NYU VAGC & -20.4 & 0.05 & 0.1 & 156266 &  5 \\ 
bright1 & NYU VAGC & -21.35 & 0.1 & 0.15 & 113713 &  8 \\ 
bright2 & NYU VAGC & -22.05 & 0.15 & 0.2 & 43340 & 13 \\ 
lrgdim & LRGs & -21.2 & 0.16 & 0.36 & 67567 & 24 \\ 
lrgbright & LRGs & -21.8 & 0.36 & 0.44 & 15212 & 38 \\ 

\hline
\end{tabular}
\label{tab:samples}
\end{table*}

%-------------------------------------------------------------------------------
\section{Voids in survey data}
\label{sec:finding}

\subsection{Coordinates}

Given a galaxy's sky latitude $\theta$, sky longitude $\phi$, and 
redshift $z$, we transform to a hybrid coordinate system
\begin{eqnarray*}
  x' & = & \frac{cz}{H_0} \cos{\phi} \cos{\theta}, \\
  y' & = & \frac{cz}{H_0} \sin{\phi} \cos{\theta}, \\
  z' & = & \frac{cz}{H_0} \sin{\theta}, 
\label{eq:transform}
\end{eqnarray*}
where $c$ is the speed of light and $H_0$ is the Hubble parameter
at redshift $z=0$.
This coordinate system preserves relative distances; thus, we are 
essentially finding voids in redshift space. This choice is motivated by
our desire to apply the Alcock-Paczynski test to void
shapes~\citep{Ryden1995} in a forthcoming work. 
For completeness, we will also make publicly
available a void catalog derived from galaxies in real space with the 
same cosmology used to construct the volume-limited samples above.
Our samples are no longer strictly volume-limited with this choice of 
coordinates; however, the void properties that we are interested in should be 
only mildly sensitive to this in the narrow range of each redshift bin.
Also, as we will discuss below the Voronoi tessellation preserves topological 
information. This means that mild 
distortions due to coordinate transformations will
not destroy voids. 

\subsection{Defining voids}
We identify voids using a substantially
modified version of the parameter-free void finder 
\zobov~\citep{Neyrinck2008,LavauxGuilhem2011},
 which is based on a Voronoi tessellation 
of the tracer particles (in this case, galaxies) to reconstruct the 
density field~\citep{VandeWeygaert2007, Platen2011} followed by a 
watershed algorithm to group Voronoi cells into zones and 
subsequently voids~\citep{Platen2007}. 

The algorithm works as follows. First, we build a Delaunay tessellation 
of the volume from the tracer positions in redshift coordinates. 
Second, we assign a density to each galaxy based on its Voronoi volume.
Using this estimated density \zobov~executes a procedure similar 
to the Watershed algorithm: the entire volume is split 
into ``zones'', with each zone corresponding to the attraction 
basin of a local minimum of the density field. 
\zobov~then assigns a core density to each zone that 
corresponds to the lowest local minimum and assembles these zones
 into voids by successively joining pairs of zones if they 
share the lowest common saddle point in the density field. 
This approach is related to the theory of persistence brought 
to the field of large scale structures by~\citet{Sousbie2011}.
Note that there are \emph{no} parameters that control the density 
determination, the construction of zones, their evolution into 
voids, or indeed any portion of this algorithm.

Under this framework a ``void'' is a collection of zones that
share common saddle points; walls, filaments, and clusters naturally 
divide a given volume into these voids. The volume of the void is 
defined as the sum of the Voronoi volumes of its zones.
 We may record these voids either 
by their geometrical definition (e.g., a collection of Voronoi cells) 
or by the tracers contained within them. 

While all available void finders accurately identify the largest 
voids in simulations, there are many differences at the smallest 
void sizes~\citep{Colberg2008}, and all potentially suffer from effects such
as artificial bridging between voids~\citep{Park2009}.
However, methods based
on Voronoi tessellation offer several advantages over competing 
methods~\citep[e.g.,][]{Aikio1998, Foster2009, Forero-Romero2009, Gaite2009,Way2011}, which typically 
gather voids by building overlapping spheres of underdensities.
First, a Voronoi tessellation provides a natural  
construction of the local density of each particle, and hence an accurate 
measurement of the void volume. 
This construction is directly related to the local number of 
neighbors of a particle: more local neighbors induce smaller 
Voronoi cells and thus a higher local density, as expected.
The use of topological criteria, like local minima and saddle points,
is what makes a void finder like \zobov~or 
{\tt DisPerSe}~\citep{Sousbie2011} particularly appealing for measuring 
geometry distortions. 
Since the topology is resilient to metric transformations, these 
methods recover mostly the same structures even if those structures have been 
affected by redshift distortions.
This resilience is essential for our analysis because we will be eliminating 
potential voids and arranging them in a tree based on density 
and volume characteristics.
Secondly, the tessellation procedure allows voids to assume arbitrary shapes
while still obeying an overall mean density threshold during the watershed
phase. 
Finally, \zobov~is able to work cleanly with any sampling of 
the density field --- from sparse LRGs in surveys to dense dark matter
particles in simulations --- without fine-tuning or parameter adjustments.

Analogous to groups and clusters, voids are naturally organized into a
tree-like hierarchy~\citep{Aragon2012, Paranjape2012}. Following the 
technique of~\citet{LavauxGuilhem2011}, we organize voids into such a 
tree based on the natural definitions of boundaries and 
basins provided by \zobov~(see Figure 1 of~\citealt{LavauxGuilhem2011}). 
The tree is built from the smallest void to the largest. 
A given 
void accepts another void as a parent in the tree if all the zones of the
void are present in the parent void and the parent void is larger than 
the considered void. This structure allows us to 
double count regions that are 
sampled by different void sizes at the same time. 
This is particularly useful when grouping void sizes into
 large bins.

We impose two independent density-based selection criteria.
First, we restrict \zobov~to report only voids with mean density contrasts
of $-0.8$ (in practice, we require the mean density contrast of the set of 
Voronoi cells that constitute a void to be $\le -0.8$; voids will generally 
have density contrasts slightly below this threshold). This reduces the effects of sampling noise in the estimated 
size of each void~\citep{Schmidt2001}.
Next we reject any void that has overdensity greater than $-0.8$ within 
a central region, which we take to be one quarter the effective 
radius (defined below). We do this to eliminate the 
effects of Poisson noise in these extremely underdense regions.
Note that these quantities are defined relative to the mean 
number density of galaxies in the survey volume. 
Finally, for a given survey sample, we ignore any voids whose 
effective radius is smaller than the mean galaxy 
separation, which analyses have shown to be the minimum resolvable 
void size~\citep{Colberg2005, Tikhonov2006, Platen2011}.

In the following discussion, we define the center of each void to be the 
volume-weighted barycenter of all the galaxies contained in the void volume:
\begin{equation}
  {\bf X}_v = \frac{1}{\sum_i V_i} \sum_i {\bf x}_i V_i,
\label{eq:barycenter}
\end{equation}
where ${\bf x}_i$ is the position and Voronoi volume of each galaxy $i$.
In other words, this is the volume-weighted average position of the 
Voronoi cells which make up a void, and also the density-weighted 
average position of the galaxies.
This reduces the effects of shot noise on the determination 
of the void center compared to other methods, such as unweighted averaging 
or choosing the most-underdense galaxy position.
Also, we define the void radius as the effective radius, or the radius 
of the sphere which has the same volume as the Voronoi-based void volume:
\begin{equation}
  R = \left( \frac{3}{4 \pi} V \right)^{1/3}.
\label{eq:effr}
\end{equation}

\begin{figure*} 
  \centering 
  {\includegraphics[type=png,ext=.png,read=.png,width=0.49\textwidth]
    {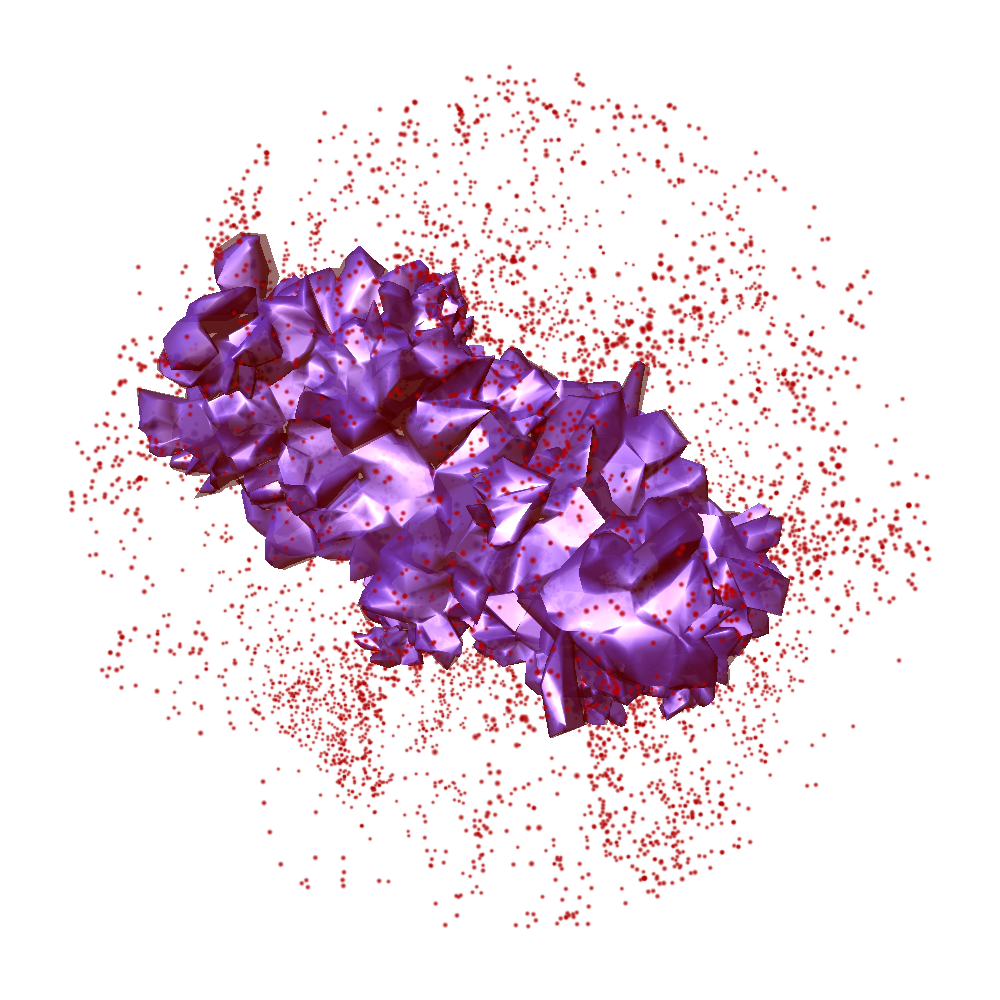}}
  {\includegraphics[type=pdf,ext=.pdf,read=.pdf,width=0.49\textwidth]
    {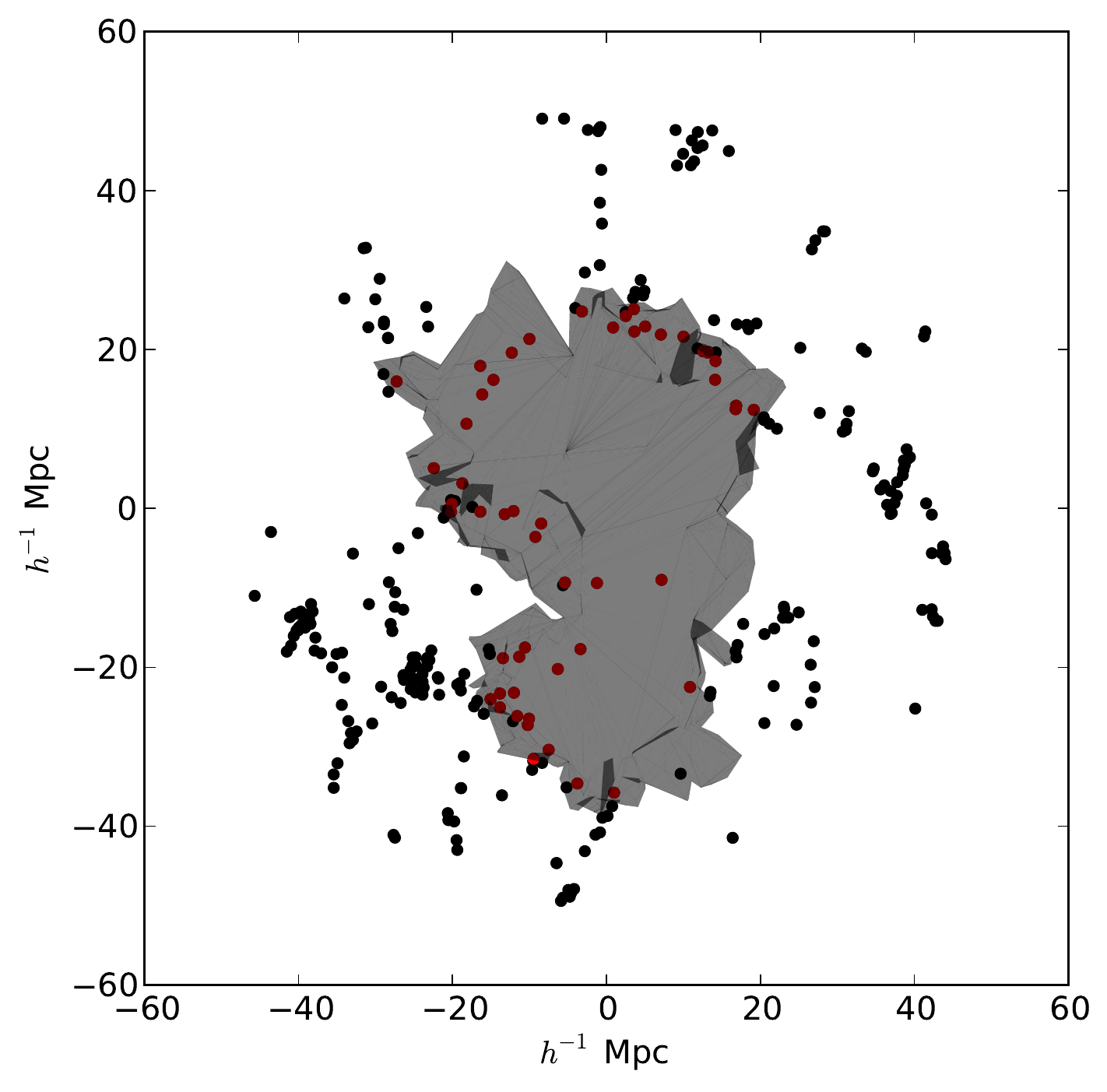}}
  \caption{\emph{Example Voronoi-based void.} 
           We show the Voronoi cells that define the void in purple
           with galaxies in red (left). Shown is a void with effective 
           radius 20~\hmpc~within 
           a spherical region of radius 50~\hmpc. 
           Galaxy point sizes are proportional 
           to their distance from the point of view. Galaxies 
           interior to the void are shaded dark red.
           On the right is a 5~\hmpc~thick slice through the same void,
           showing exterior galaxies in black and interior galaxies in red.
           The orientation of the void is different between the panels 
           to highlight different aspects of the structure.
           }
\label{fig:void}
\end{figure*}

Figure~\ref{fig:void} shows an example $R=20$~\hmpc~void 
from the \emph{dim2} sample with both its galaxy members and 
surrounding non-members. This highlights how the Voronoi and 
watershed technique 
of \zobov~stretches and deforms the void shape to fill the entire 
underdense region. The void is clearly buttressed by denser patches 
of galaxies identified as clusters and walls.
The highly non-spherical shape is characteristic of voids identified 
with the Watershed Transform. One consequence, which will become apparent 
below, is that a \emph{sphere} of radius $R$ is not necessarily underdense, 
even though the Voronoi volume itself has a mean density contrast of $-0.8$.

\subsection{Accounting for survey boundaries}

\begin{figure} 
  \centering 
  {\includegraphics[width=\columnwidth]{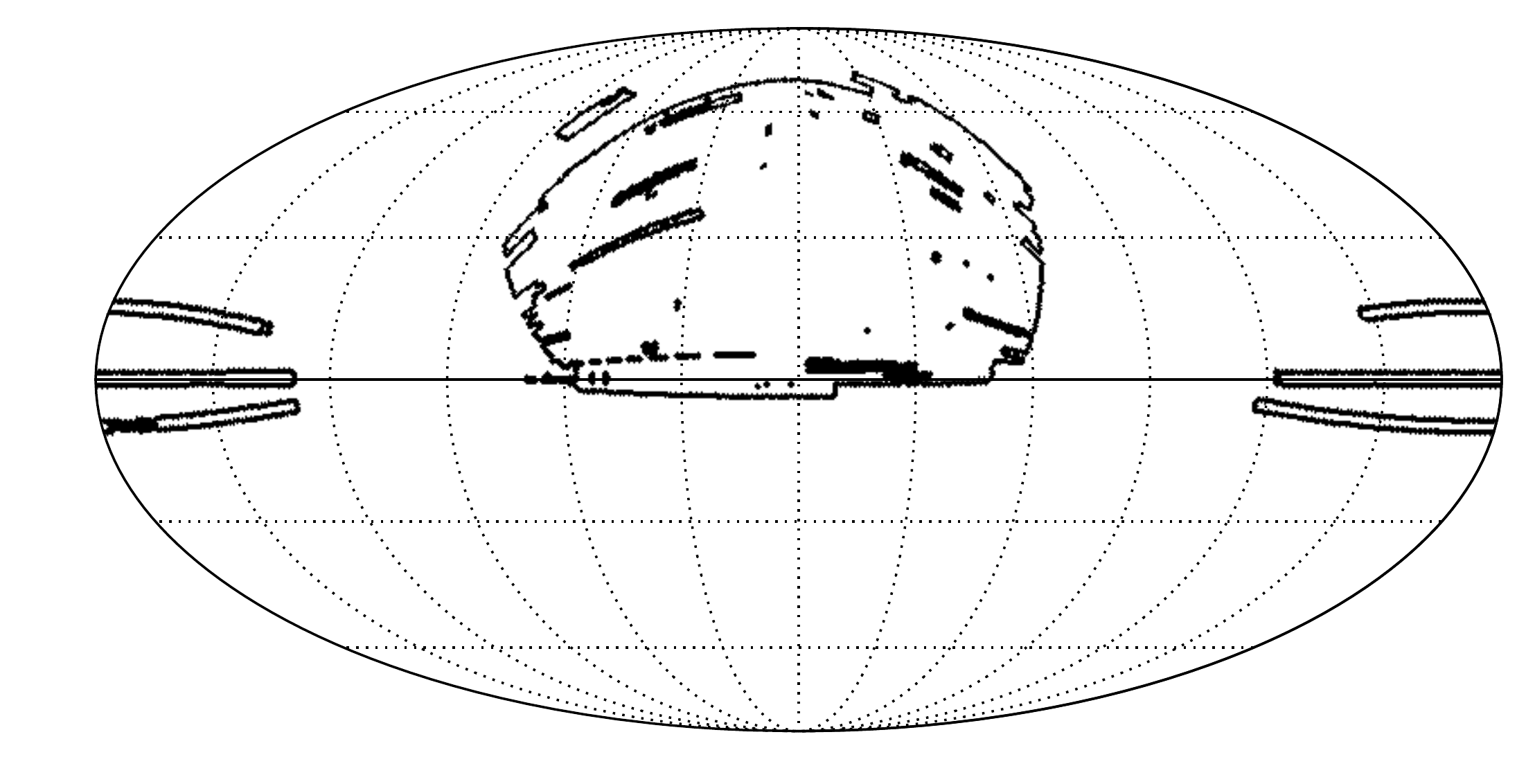}}
  \caption{\emph{Using \healpix~to identify boundaries.}
           %Left: \healpix~map of SDSS mask. Black indicates survey area.
           \healpix~map in a Mollweide projection 
                  of identified boundary zones (black) around and  
                  within the SDSS survey area where 
                  we inject boundary particles.  
           }
\label{fig:mask}
\end{figure}

We must make a few modification to \zobov~to account for survey masks, 
boundaries, and redshift cutoffs so that voids do not include 
volumes outside the survey extent.
We begin with a pixelization of the survey mask 
using \healpix~\citep{Gorski2005}\footnote{\url{http://healpix.jpl.nasa.gov}}. 
The \healpix~description of the sphere provides
equal-area pixels, and the \healpix~implementation 
itself provides built-in tools to easily
determine which pixels lie
on the boundary between the survey area and any masked region.
Figure~\ref{fig:mask} shows our pixelization of the SDSS DR7 mask
and the location of the boundary pixels. To accurately capture 
the shape of the mask we required a resolution of $n_{\rm side}=512$.

We wish to prevent voids from extending into the boundary 
or near any masks within the survey area. 
To accomplish this
we inject boundary particles along each boundary pixel with a 
density $10^{-3} (h^{-1} \rm {Mpc})^{-3}$.
Testing has shown that we require an order of magnitude lower density 
to ensure that every galaxy near the survey edge is closer to a boundary 
particle than any other galaxy and to stabilize the resulting voids 
for the most dense sample; lower density 
samples require even fewer boundary particles.
We place the boundary particles randomly within the volume defined by the 
surface area of the \healpix~pixel and the redshift extent along the line 
of sight. 
This process and our chosen density 
 results in a very thin sheath of particles that completely encloses 
the galaxy distribution.
In addition, we place boundary particles at the minimum and maximum redshift 
``caps'' of the survey. Finally, we define a cubic box that completely 
contains the survey volume and distribute boundary particles evenly along the 
surface of that box. This last placement provides closure to the 
Voronoi tessellations of the boundary particles so that we do not have to 
directly modify this portion of \zobov.

We assign infinite densities to the boundary particles --- this prevents 
boundary particles from joining zones and voids and hence stops voids 
from expanding past the survey region.
In addition, since the volume of any 
Voronoi cell that touches a boundary particle is by definition 
arbitrary, we cannot include that cell in any constructed void. 
Thus, after the tessellation phase we remove both the boundary particles and 
any galaxies adjacent to the boundary population (i.e., closer to a 
boundary particle than any other galaxy) together with their associated 
Voronoi cells. 
This process removed 
approximately 10\% of the true galaxy 
population. 

\subsection{Handling mask-induced bias}

The above prescription produces a complete and robust void catalog that 
fully accounts for survey boundaries and masks: 
the Voronoi volumes that define each void lie completely within the 
survey volume, and the mean density contrast of the summed Voronoi 
volumes of each void is $\le 0.8$.
 The resulting 
catalog is appropriate for many applications, such as studying 
the fractional volume of the universe occupied by voids or studying 
the properties of galaxies within voids.

However, the survey mask, boundaries, and redshift cutoffs
introduce a subtle bias: they will 
preferentially select voids that lie along the line of sight 
(for boundaries and masks) and perpendicular to the line of site 
(for redshift cutoffs). For example, a void parallel and near to the 
survey boundary will be relatively complete, but a void perpendicular 
to the survey boundary will appear as a smaller, truncated void, 
which will tend to be below the resolution threshold and hence 
expunged from the catalog.
Thus, analyses that rely on the \emph{shapes} of voids, 
such as the ellipticity probability distribution or the 
Alcock-Paczynski test, will be strongly affected because
they will see more voids aligned with the 
mask and boundary than perpendicular to it.

To eliminate this bias, we produce a culled sample of voids 
that we refer to as the ``central'' sample in later discussions. 
To produce this sample we remove any void which, when
 rotated in any direction about its 
barycenter, intersects a boundary galaxy. 
Operationally we perform this by taking the distance from the 
void barycenter to the furthermost particle and comparing that 
to the distance to the nearest boundary particle.
This ensures that we have a fair distribution of void shapes and 
alignments within the survey volume.

%-------------------------------------------------------------------------------
\section{Void demographics}
\label{sec:properties}

\begin{figure*} 
  \centering 
  {\includegraphics[type=png,ext=.png,read=.png,width=0.75\textwidth]
    {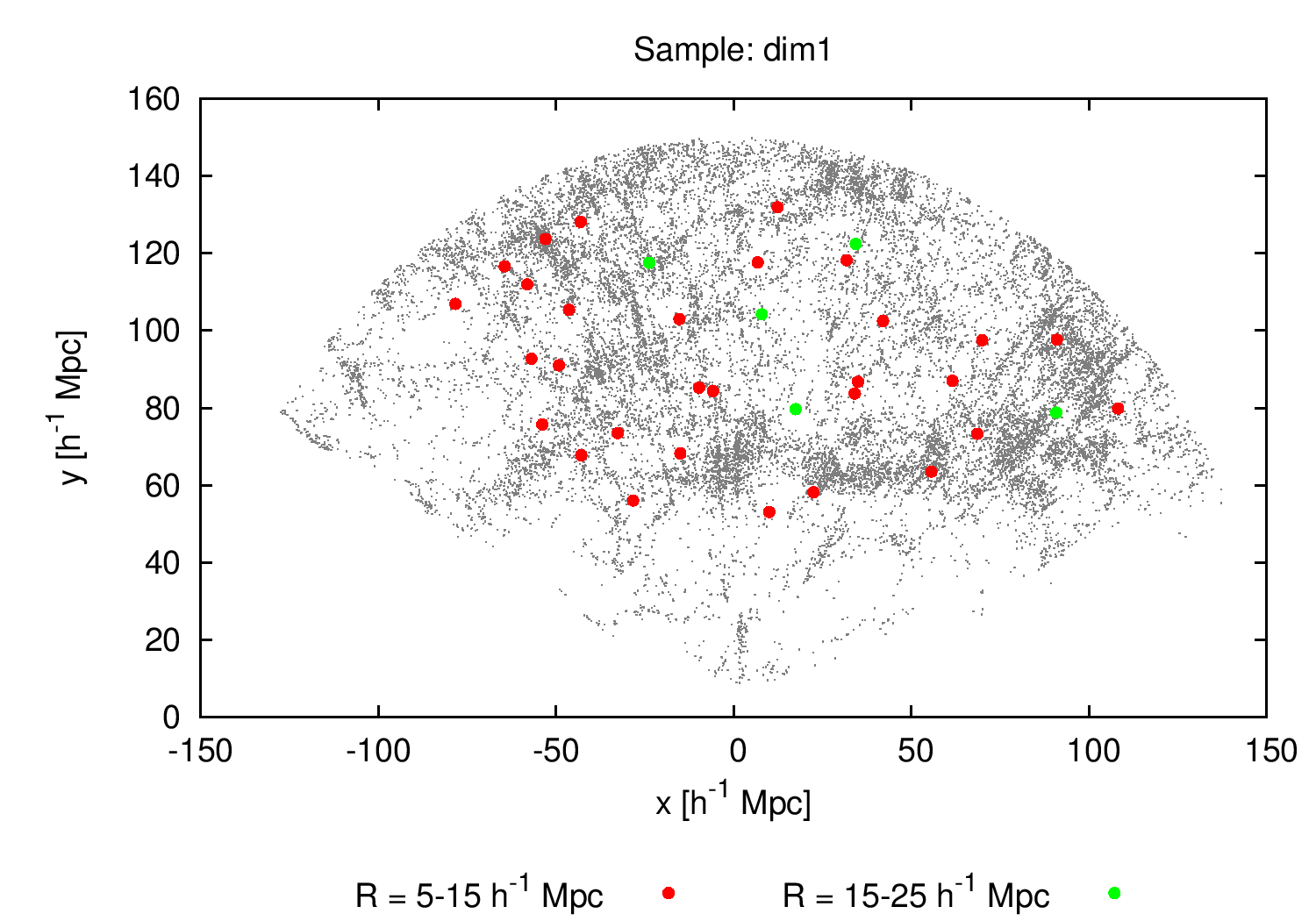}}
  {\includegraphics[type=png,ext=.png,read=.png,width=0.75\textwidth]
    {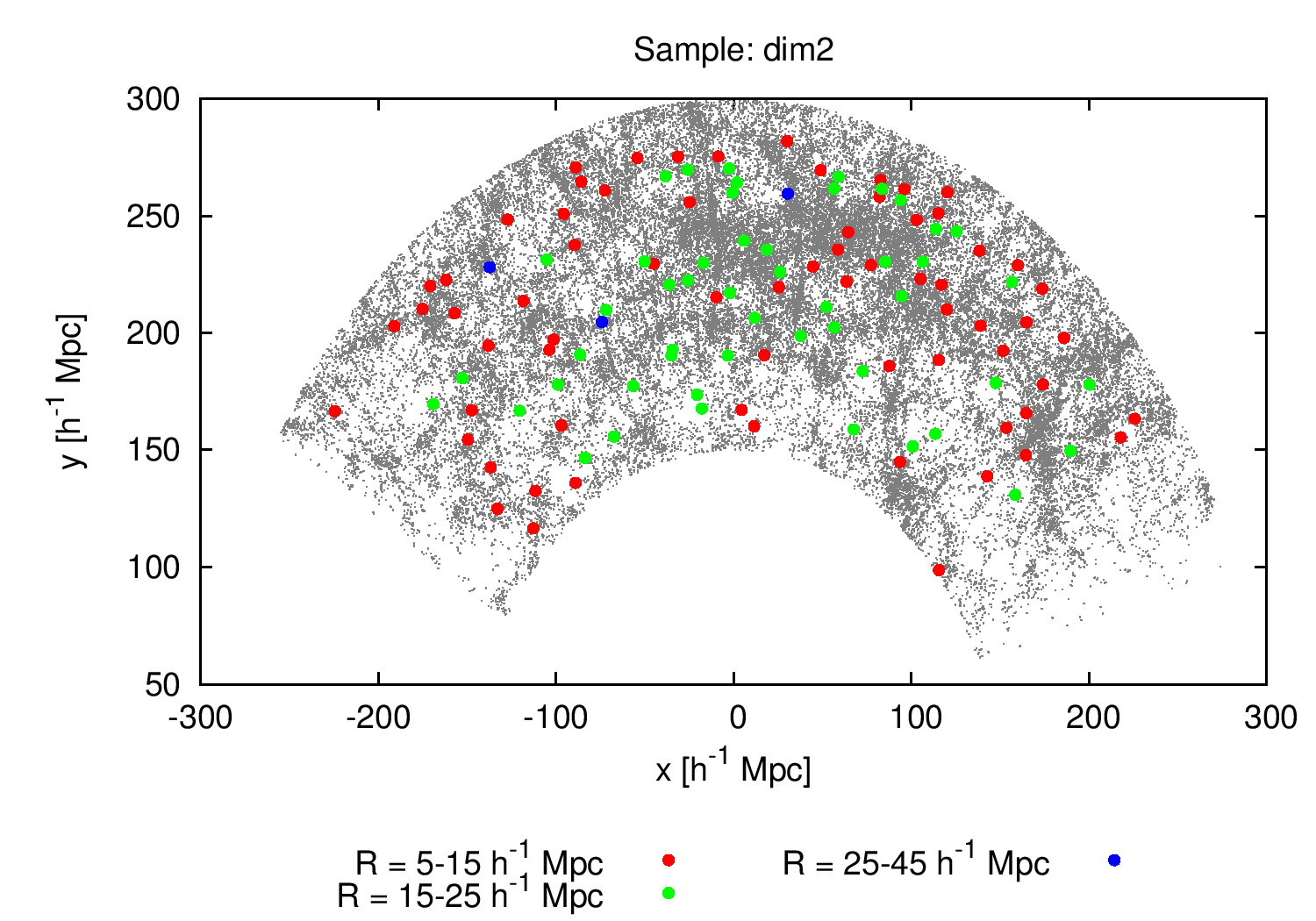}}
  %{\includegraphics[type=png,ext=.png,read=.png,width=0.9\textwidth]
  %  {f4a}}
  %{\includegraphics[type=png,ext=.png,read=.png,width=0.9\textwidth]
  %  {f4b}}
  \caption{\emph{Void distributions throughout each \emph{dim} sample.} 
           We show the spatial distribution of all voids for 
           the \emph{dim1} and \emph{dim2}  volume-limited samples. 
           We plot galaxies in grey and void barycenters in colors depending 
           on the size as indicated in the plots. We compact the galaxy 
           and void locations by rotating their positions about the 
           $x$-axis to lie
           within the $x$-$y$ plane, and only show galaxies and voids within 
           a $25$ degree opening angle through the survey volume.}
\label{fig:locations1}
\end{figure*}

\begin{figure*} 
  \centering 
  {\includegraphics[type=png,ext=.png,read=.png,width=0.9\textwidth]
    {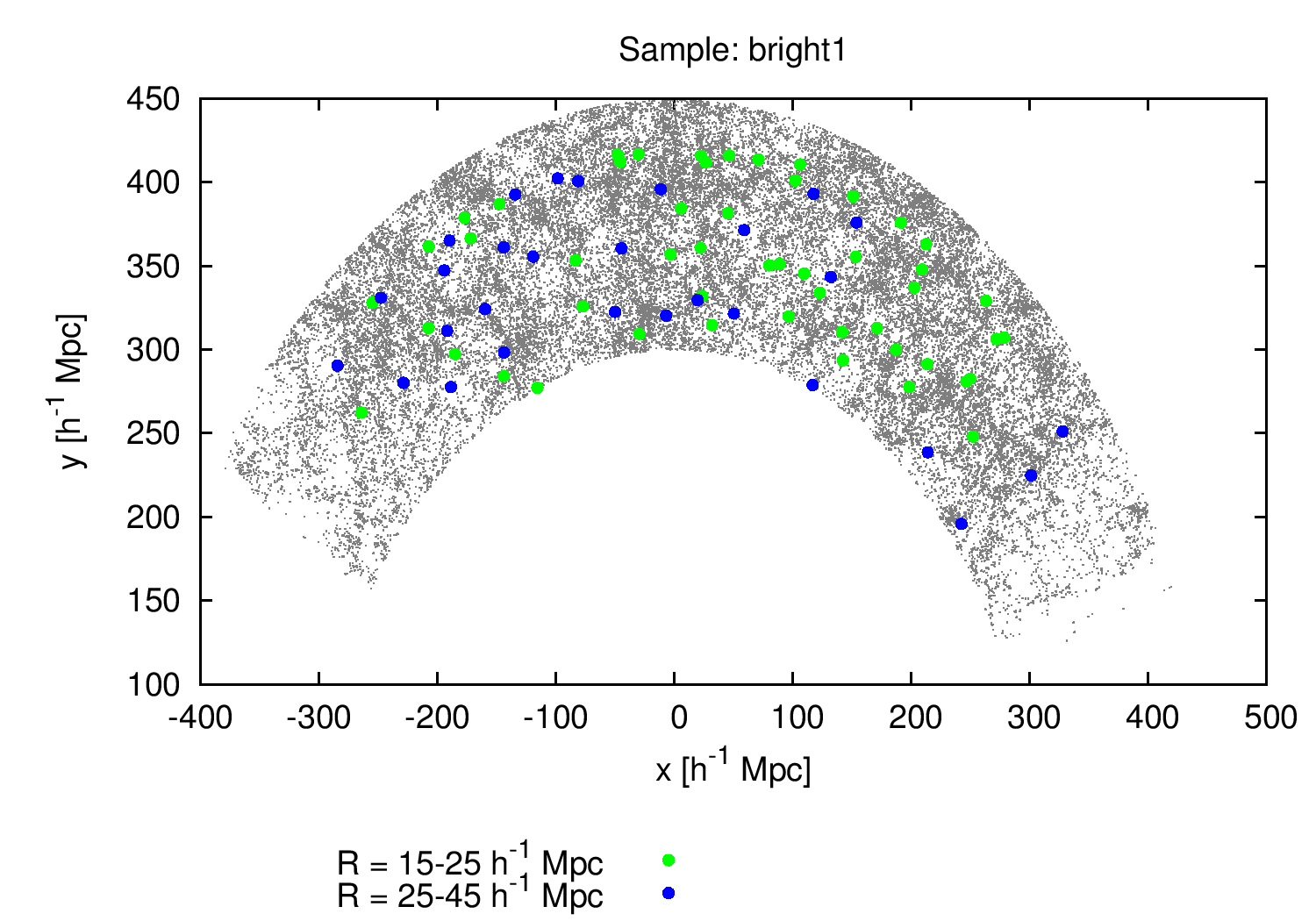}}
  {\includegraphics[type=png,ext=.png,read=.png,width=0.9\textwidth]
    {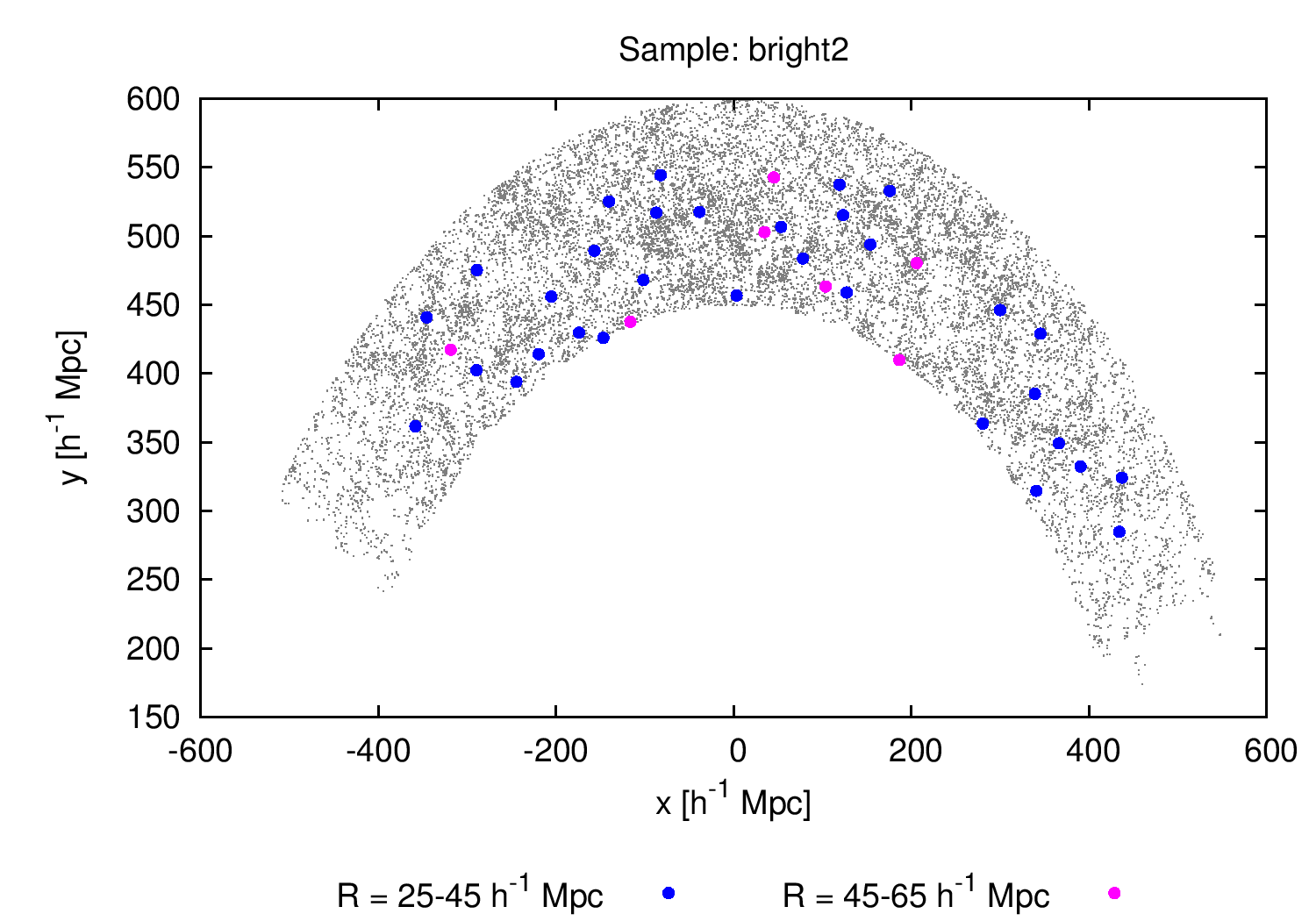}}
  \caption{\emph{Void distributions throughout each \emph{bright} sample.} 
           We show the spatial distribution of all voids for 
           the \emph{bright1} and \emph{bright2}  volume-limited samples. 
           See Figure~\ref{fig:locations1} for a plot description.
           }
\label{fig:locations2}
\end{figure*}

\begin{figure*} 
  \centering 
  {\includegraphics[type=png,ext=.png,read=.png,width=0.9\textwidth]
    {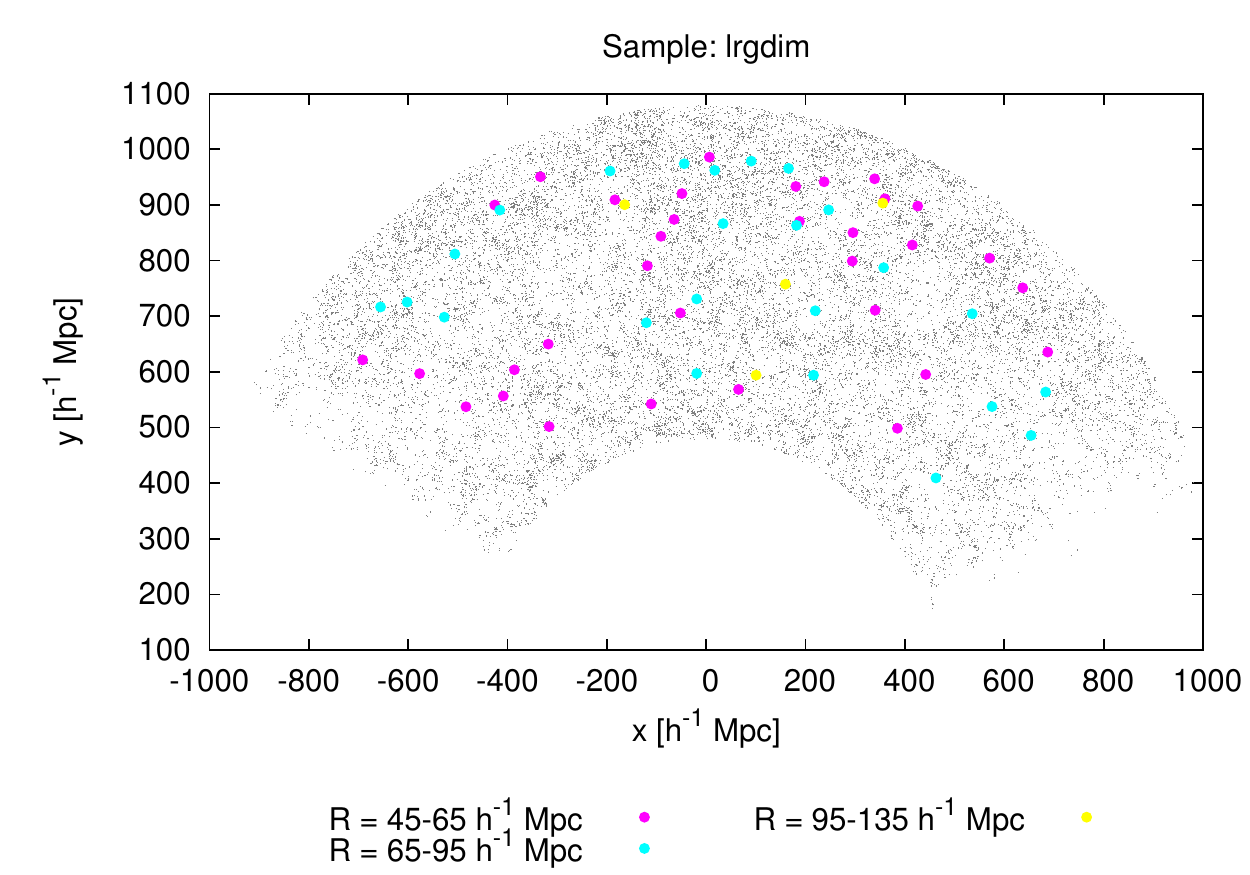}}
  {\includegraphics[type=png,ext=.png,read=.png,width=0.9\textwidth]
    {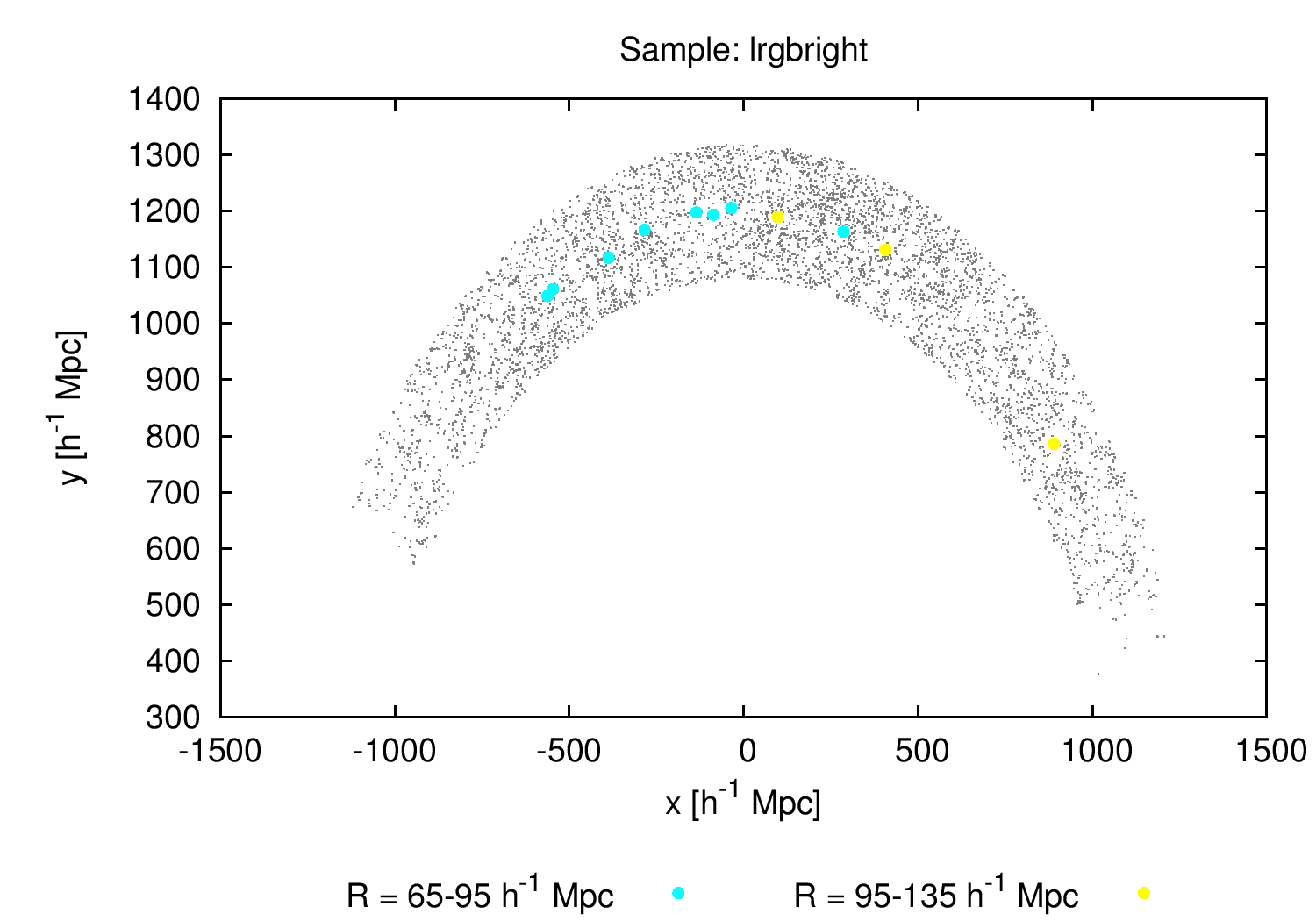}}
  \caption{\emph{Void distributions throughout each \emph{lrg} sample.} 
           We show the spatial distribution of all voids for 
           the \emph{lrgdim} and \emph{lrgbright} volume-limited samples. 
           See Figure~\ref{fig:locations1} for a plot description.
           }
\label{fig:locations3}
\end{figure*}

We begin with a discussion of the identified void locations and distributions 
within the four volume-limited samples.
Figures~\ref{fig:locations1},~\ref{fig:locations2}, and~\ref{fig:locations3}
 show the barycenters of all voids 
in each sample overlaid on galaxy positions. To clarify plotting, 
we have rotated all galaxy 
and void positions about the $x$-axis so that they lie on the $x$-$y$ plane
(i.e., a rotational projection). We only show galaxies and voids within 
a 25~degree opening angle.
We plot the void positions with different colors to distinguish 
their sizes.

We immediately notice that voids naturally avoid the edges of the survey volume, with the exception of the lower-redshift boundary, where high-redshift samples 
include low-redshift galaxies in order to capture as many voids near the 
boundaries as possible.
Voids that intersect the edges tend to be smaller, and we remove 
from the catalog any voids smaller than the mean galaxy 
separation. 
This is especially evident in the \emph{lrgbright} subsample:
since the mean galaxy separation is so large compared to the redshift 
extent of the subsample, we only find the few voids near the 
median redshift. 
Despite this natural edge-avoidance, 
as we will see below there are still voids that need 
to be removed to produce a fully bias-free sample. We do not find any 
voids in any sample in the southern sky; the 
survey slices are so thin that they truncate any identified void there.

The surviving voids distribute as expected: we clearly see smaller 
voids ``hugging'' the edges of filaments and sheets, while larger voids 
inhabit the more expansive empty regions of the survey. As we move to 
higher-redshift and sparser samples, we see ever-larger void sizes,
although for each sample the same general distribution of smaller 
and larger voids applies. In all samples, we tend to find more small
voids near the boundaries, since the mask tends to miss small voids while 
truncating larger voids, making them appear smaller.

\begin{figure} 
  \centering 
  {\includegraphics[type=pdf,ext=.pdf,read=.pdf,width=\columnwidth]
    {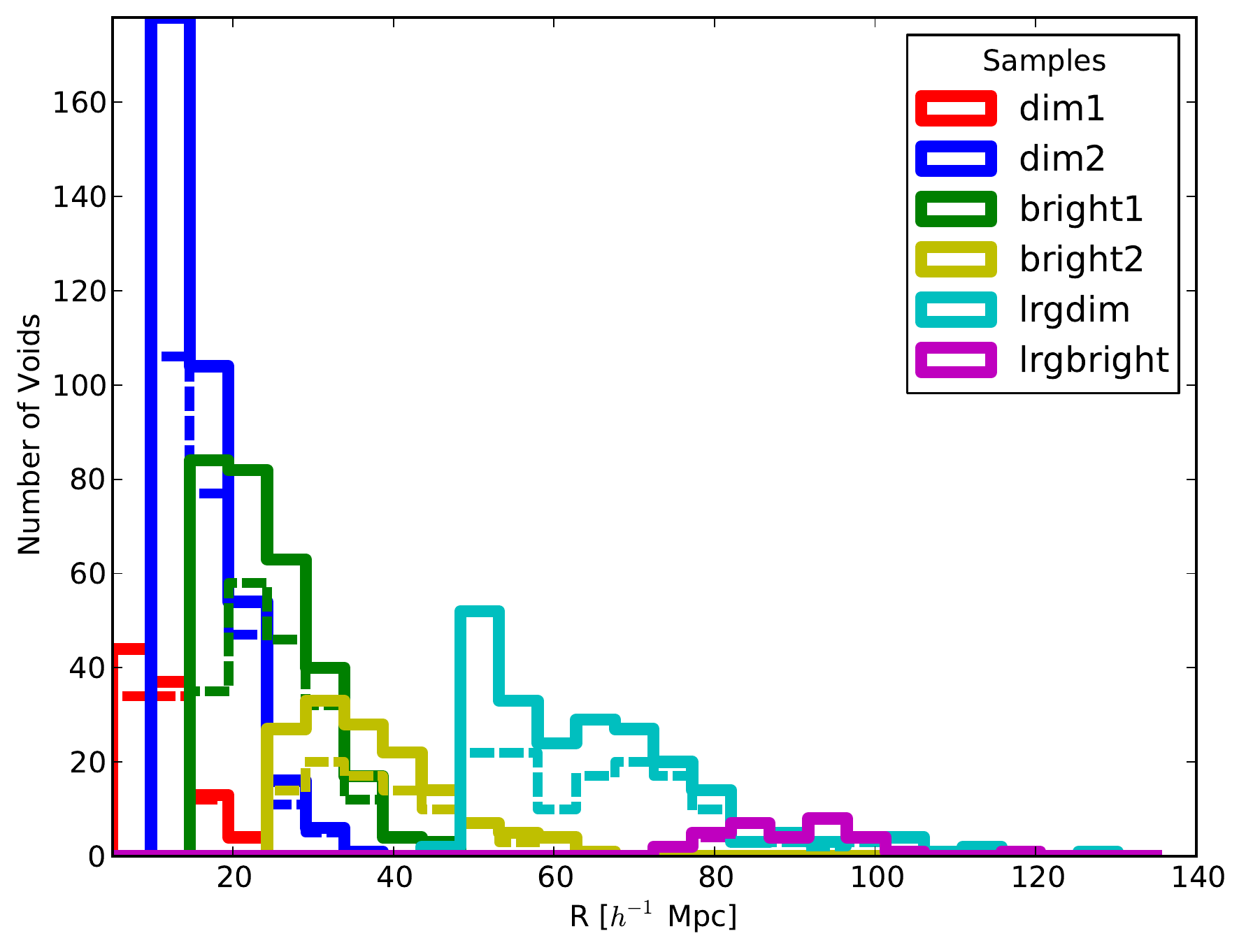}}
  \caption{\emph{Distribution of void sizes.}
           We plot histograms of void radii, colored by sample.
           Solid lines are from the 
           all-void sample, while dashed lines are central voids.}
\label{fig:rhist_allsamples}
\end{figure}

Figure~\ref{fig:rhist_allsamples} shows the size distribution of 
voids in each of our samples. 
While we see the same number of large voids in the \emph{dim1} and 
\emph{dim2} samples
as in the study of 
~\citet{Pan2011}, which used a single volume-limited sample out to 
$z=0.1$, we see many smaller voids because we are using an extra 
low-redshift subsample. 
Each successively higher-redshift sample produces larger voids. 
Our smallest voids, in the \emph{dim1} sample, have effective 
radii $\sim 5$~\hmpc, while the very largest in the \emph{lrgdim} 
reach $R = 135$~\hmpc.
This is caused by several factors. First, brighter galaxies 
are more strongly biased, which can lead to larger voids at 
higher redshifts.
Second, shot noise is larger in a sparser sample, and some void 
sizes may be enhanced by including regions that 
are underdense because of sampling effects, but~\citet{Bos2012} find
that this is a small effect.  
Third, at lower redshifts the ratio of survey surface area to 
volume is much higher; this truncates any potential large 
void at low redshift. Finally, there is a simple volume effect: 
if a void of a given size has a given probability to 
occupy a region of space due to cosmic variance, we require a 
certain minimum sample volume to discover it.
While there is some overlap of void sizes between adjacent samples, 
the multiple samples allow us to extract a large range 
of void sizes.

By removing voids that can intersect the mask after rotation,
 we mostly affect the smaller voids 
of each sample. The mask already 
truncates larger voids so that they tend to appear 
as smaller voids. Thus, the remaining large voids are unaffected during 
the production of the central catalog. This is especially true for the 
\emph{lrgbright} sample, where the catalog contains only the very largest 
voids, which all survive in the central catalog. 
The de-biasing procedure removes 
$30-50\%$ of the smaller voids in each sample.
 
\begin{figure} 
  \centering 
  \includegraphics[type=pdf,ext=.pdf,read=.pdf,width=\columnwidth]
    {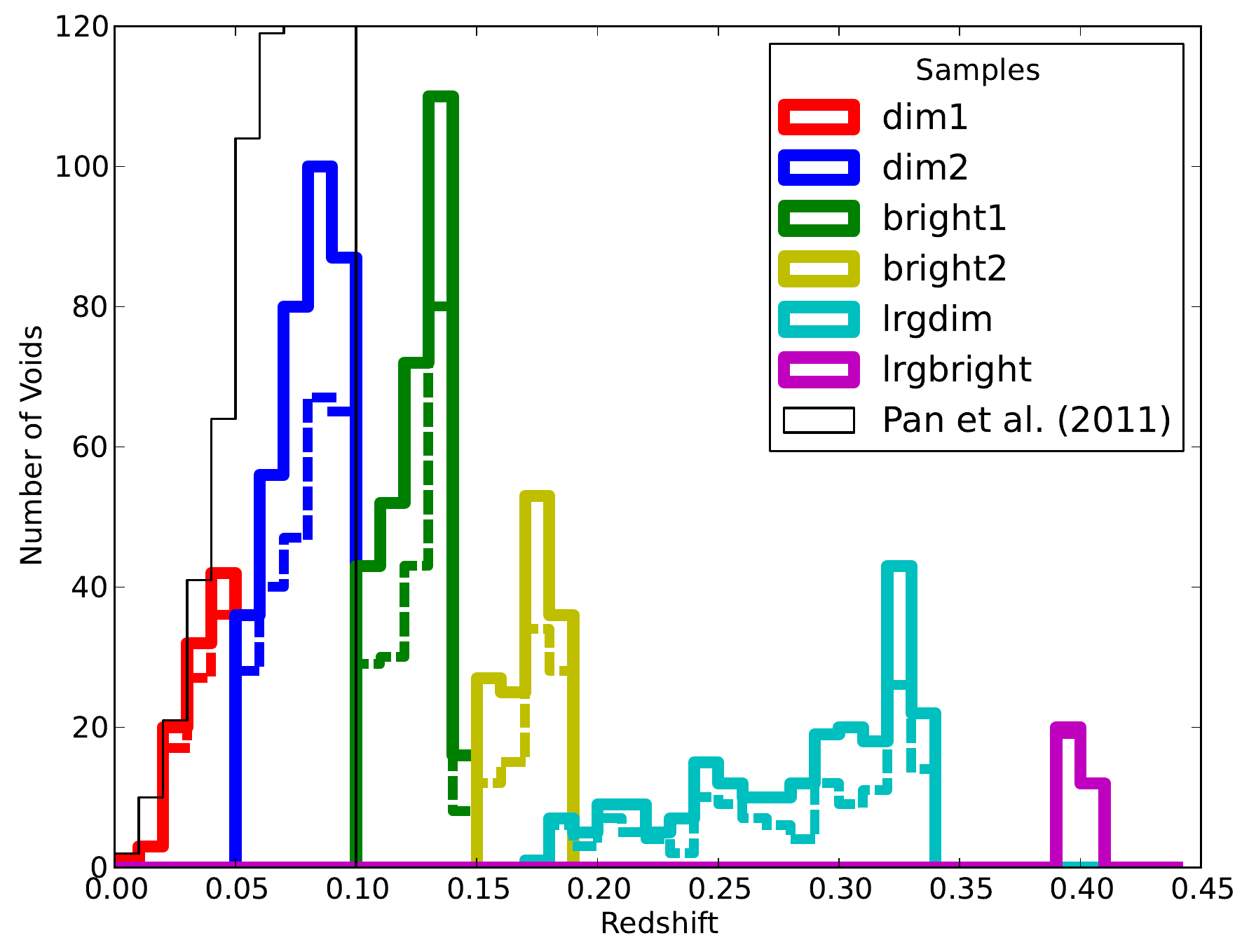}
  \caption{\emph{Redshift-dependent number counts.}
           We show binned number counts as a function of redshift, 
           colored by sample.
           Number counts of all void sizes 
           as a function of redshift for each sample show that 
           small voids are much more common in the data. Solid 
           lines are from the all-void catalog; dashed lines are 
           central voids.
           The thin black line is the redshift distribution of voids in the 
           ~\citet{Pan2011} void catalog.}
\label{fig:numbercounts}
\end{figure}

We show binned void number counts as a function of redshift 
in Figure~\ref{fig:numbercounts}.
Below redshift 0.05 we roughly agree with the number counts from the 
~\citet{Pan2011} void catalog, except at the redshift cap of 
the \emph{dim1} sample, where we remove voids near the edge.
In the \emph{dim2} sample, we count roughly half as many voids as 
~\citet{Pan2011}.
This is not surprising since we have several strict void criteria: 
a minimum void size, a maximum density threshold, and a maximum central 
density threshold. The last criterion removes a significant number of voids, 
especially in lower-density samples. Thus we include many fewer voids in our 
catalog, by construction. 

Within each volume-limited sample with redshift $z<0.2$, the 
number of voids grows strongly with redshift. 
Since the volume-limited 
samples have a fixed number density of galaxies throughout the 
redshift range, we expect to be able to count all the smaller 
--- and more 
common --- voids as we probe to higher redshifts within the samples. 
This trend is not as strong for the LRG-based samples, since we can 
only identify very large voids.
For the \emph{lrgbright} sample, our voids are confined to a very 
narrow redshift range centered on $z \sim 0.4$.
Within each sample the number density per unit volume remains 
relatively constant. However, between slices the number density 
drops as we lose the smallest voids due to the reduced galaxy 
number densities.

We see the strongest effect of cutting out biased voids at the redshift caps 
of each sample: we remove $30-40\%$ of the voids at the highest 
redshift ranges of each sample and only a small fraction of the 
voids within the bulk of the sample volume. 
The surface area of each redshift boundary is much larger than the 
line-of-sight boundaries; thus, we expect to remove more truncated 
voids here.

Table~\ref{tab:volumes} summarizes the void volumes versus the available 
volume for each sample, where we list the sample volume, void fraction, 
central sample volume, and central void fraction. We calculate the sample volume by including any regions of the survey mask. Since our void finding 
approach builds voids to the very edge of the survey, this gives us the 
effective usable sample volume. The void fraction is the ratio of the total volume of all voids in that sample to the usable sample volume. The central sample
volume is an estimate of the usable volume after we have removed voids 
near the edge, and the central void fraction is the ratio of 
the total volume of voids 
in the central sample of voids to the central sample volume.

\begin{table*}
\centering
\caption{Volumes of Surveys and Voids.}
\footnotesize
\tabcolsep=0.11cm
\begin{tabular}{cccccc}
  Sample & Volume ($10^6$ \hmpccub) & Void Fraction & 
              Central Volume ($10^6$ \hmpccub) & Central Void Fraction \\
  \hline
  \hline
  dim1 & 2.7 & 0.30 & 1.9 & 0.39 & \\ 
dim2 & 18.8 & 0.43 & 15.6 & 0.40 & \\ 
bright1 & 50.9 & 0.42 & 43.3 & 0.37 & \\ 
bright2 & 99.2 & 0.37 & 74.1 & 0.34 & \\ 
lrgdim & 912.7 & 0.34 & 698.2 & 0.30 & \\ 
lrgbright & 826.2 & 0.12 & 426.9 & 0.23 & \\ 

\hline
\end{tabular}
\label{tab:volumes}
\end{table*}

%-------------------------------------------------------------------------------
\section{Radial profiles}
\label{sec:radial}

\begin{figure*} 
  \centering 
  {\includegraphics[type=pdf,ext=.pdf,read=.pdf,width=0.49\textwidth]
    {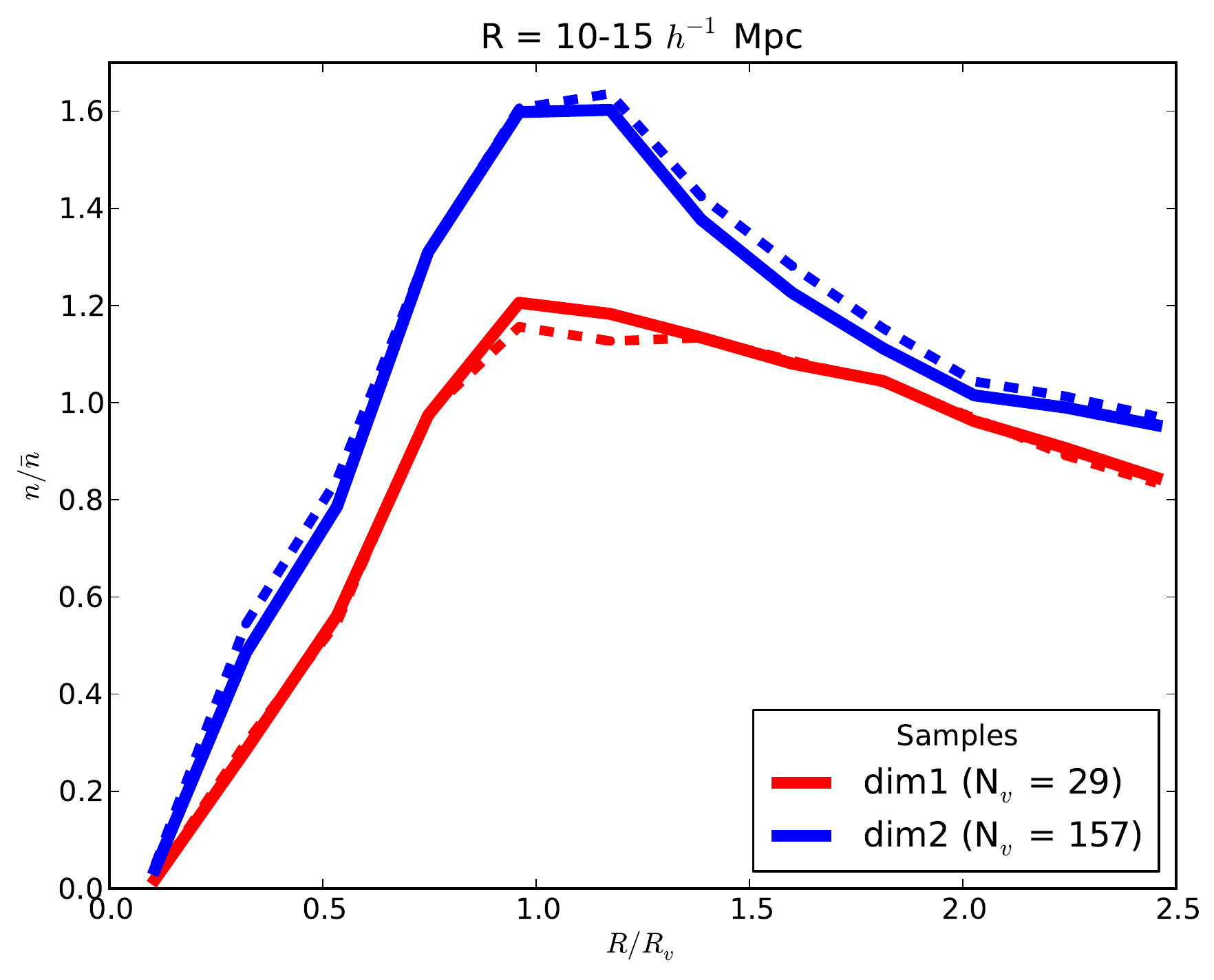}}
  {\includegraphics[type=pdf,ext=.pdf,read=.pdf,width=0.49\textwidth]
    {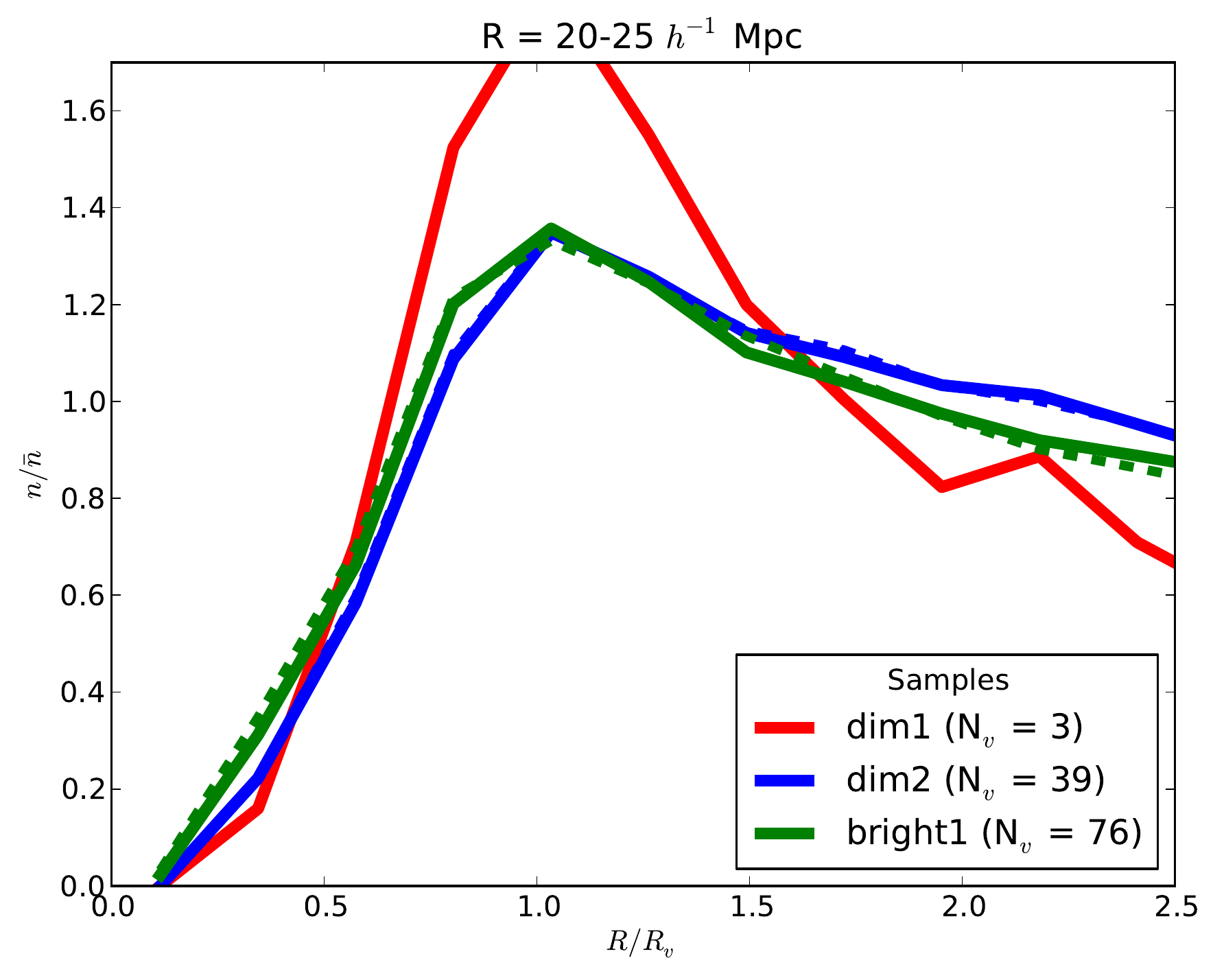}}
  {\includegraphics[type=pdf,ext=.pdf,read=.pdf,width=0.49\textwidth]
    {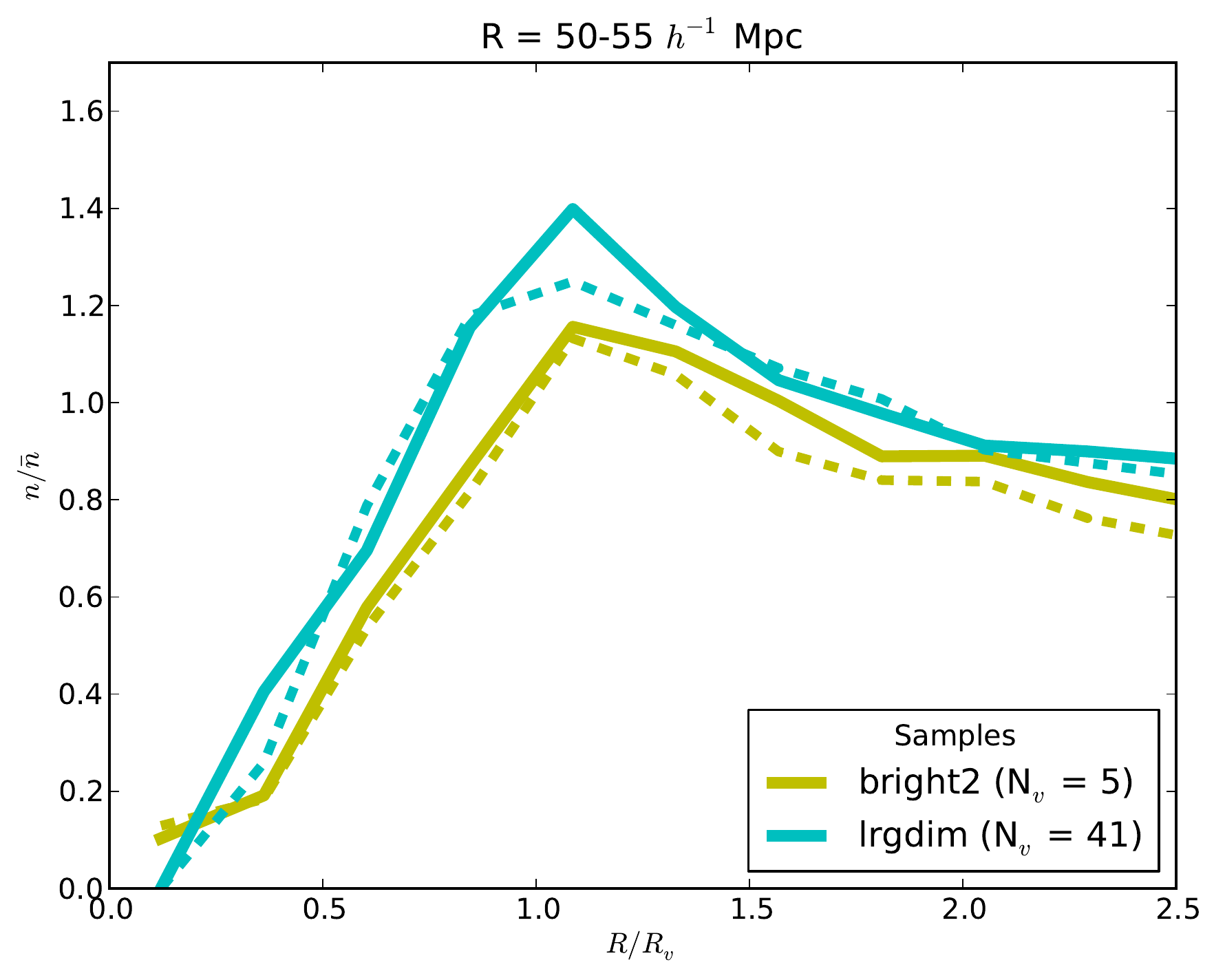}}
  {\includegraphics[type=pdf,ext=.pdf,read=.pdf,width=0.49\textwidth]
    {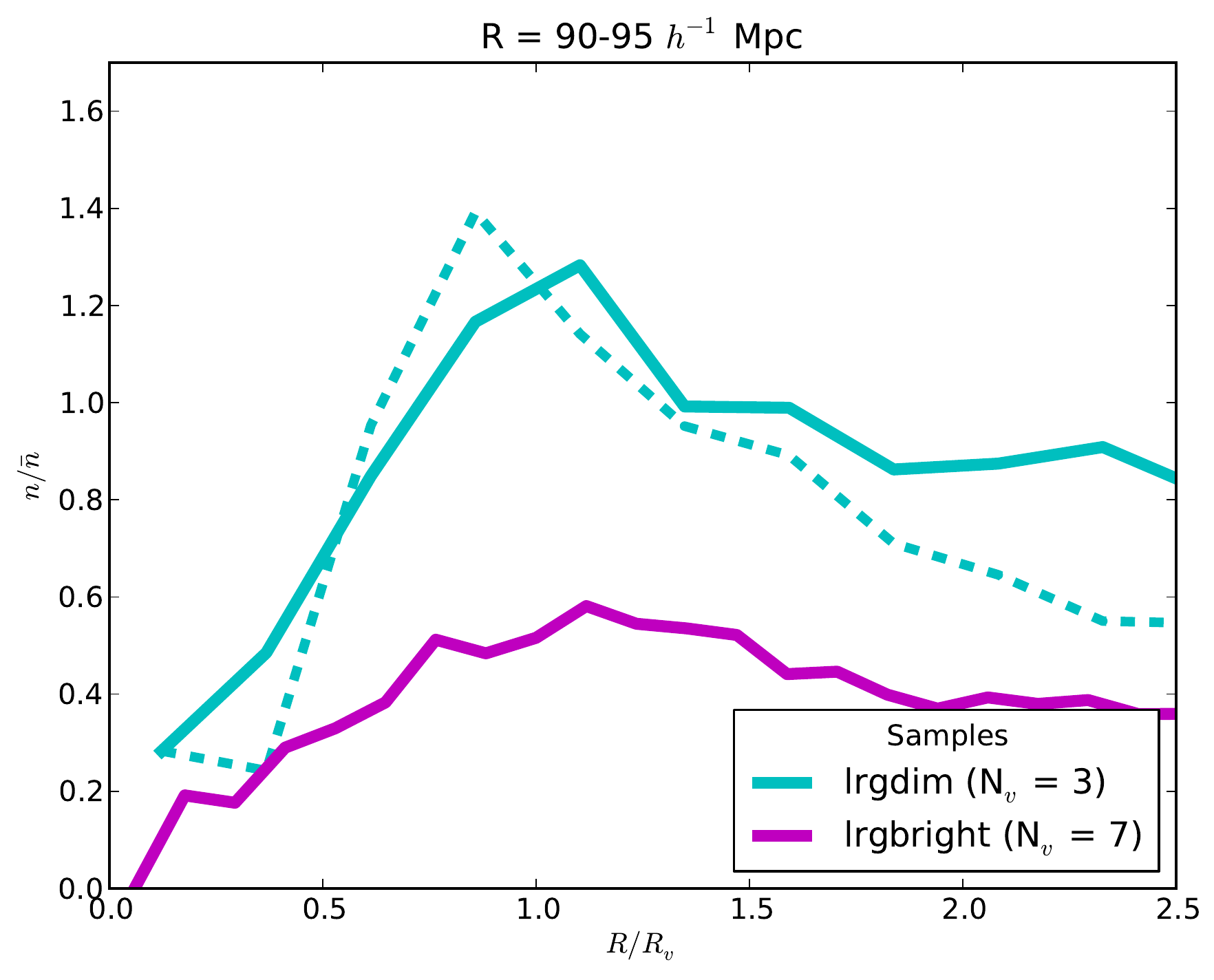}}
  \caption{\emph{Qualitatively similar radial profiles of stacked voids.} 
           Radial profiles (i.e., mean density in thin spherical shells)
            of stacked voids for various void sizes 
           versus the mean void radius in each stack
           indicate steep profiles and large compensations across all 
           void sizes. 
           Note that since different samples probe different 
           redshift ranges, these plots also give the redshift evolution 
           of similarly-sized voids (albeit identified in different types of 
           galaxies).
           We indicate the number of voids in each stack with the sample 
           name in each figure. 
           Solid lines are from the all-void catalog; 
           dashed lines are central voids.}
\label{fig:profile_samples}
\end{figure*}

We now move to another example data product of our catalog: 
radial profiles.
We build radial profiles of the mean density in thin spherical shells 
of several stacks of voids. Each stack 
contains voids with radii within a 5~\hmpc~range. To construct the stacks, 
we reposition each void such that their barycenters coincide and 
rotate them so that their directions along the line of sight 
share a common vector. We show these 
profiles in Figure~\ref{fig:profile_samples}. In this figure, we show 
stacks from the all-void catalog 
in solid lines and stacks from central voids with 
dotted lines. We also show the number of voids in the given stack.
To construct these, we do \emph{not} rescale the voids, since the 
radial bin widths are small compared to the sizes of the voids, and 
rescaling voids requires difficult calibration of the 
normalization. We have 
normalized each density profile to the mean number densities of galaxies 
in the sample, $\bar{n}$. 

These stacked void profiles clearly show a qualitatively 
similar behavior across 
all void sizes: an extremely underdense center (by construction), 
a steep wall, a large compensation at the wall, 
and a gradual decline to the mean 
density. 
These radial profiles reflect the qualitatively universal behavior of 
densities inside the shell region seen in~\citet{LavauxGuilhem2011}.
The steep walls observed here are expected in line with other 
observations: we are building profiles based on a relatively sparse 
sampling of the underdensity, 
so we expect voids from observations to have sharply-defined 
edges and large gradients at the 
walls~\citep{Fillmore1984, Benson2003, Furlanetto2006}.
Indeed, our profiles are similar to those found in previous 
works based on earlier observations and mock galaxy 
catalogs~\citep[e.g.][]{Hoyle2004, Padilla2005}.
The density profile shown in~\citet{Pan2011} does not show the 
characteristic overdensity because their voids are based on 
overlapping spherical underdensities and they average together and rescale 
voids of \emph{all} sizes.
Simulations, 
which probe the underlying dark matter distribution with high sampling density, 
produce voids with shallower walls 
\citep[e.g.,][]{Colberg2005, LavauxGuilhem2011}.

In some cases the density profile drops below the mean density at 
large distances, especially for the largest voids in the stack. 
This is especially apparent in the \emph{lrgbright} sample,
which never reaches the mean density (see the lower panel 
of Figure~\ref{fig:locations3}).
While we have constructed our catalog such that the \emph{interiors} 
of voids do not include any regions outside the survey,
at large radial distances from the void 
center we eventually encounter the mask, where the spherically-averaged 
mean density drops precipitously.
However, this is only a minor issue affecting the largest 
radii for all the other stacks.

Even though we construct our voids to have mean overdensities of 
$-0.8$, the radial profiles reach the mean density at much smaller 
radii. Since voids are elliptical, with a mean ratio of major to minor 
axis of two~\citep{Lavaux2010}, 
we expect spherical profiles to reach the void wall 
along the minor axis first, producing a steeper profile. 
Also, since we are stacking voids of different sizes without rescaling, 
the smaller voids will add to the density at smaller 
radii, again producing steeper profiles.

The effects of shot noise are apparent in stacks that 
contain few voids. For example, the $50-55$~\hmpc~
stack of \emph{bright2} voids contains only five voids, and a 
single void with an excess of galaxies near the center drives up the 
stacked density profile in that region. 

%-------------------------------------------------------------------------------
\section{Projections}
\label{sec:projected}

\begin{figure*} 
  \centering 
  {\includegraphics[type=png,ext=.png,read=.png,width=0.49\textwidth]
    {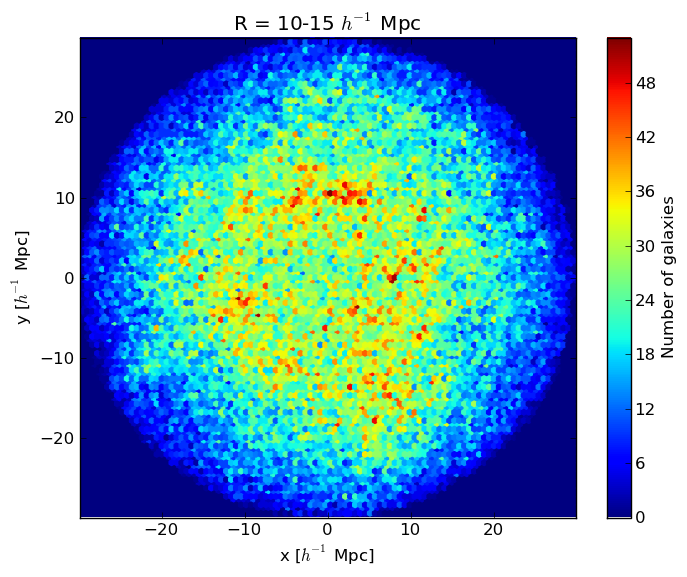}}
  {\includegraphics[type=png,ext=.png,read=.png,width=0.49\textwidth]
    {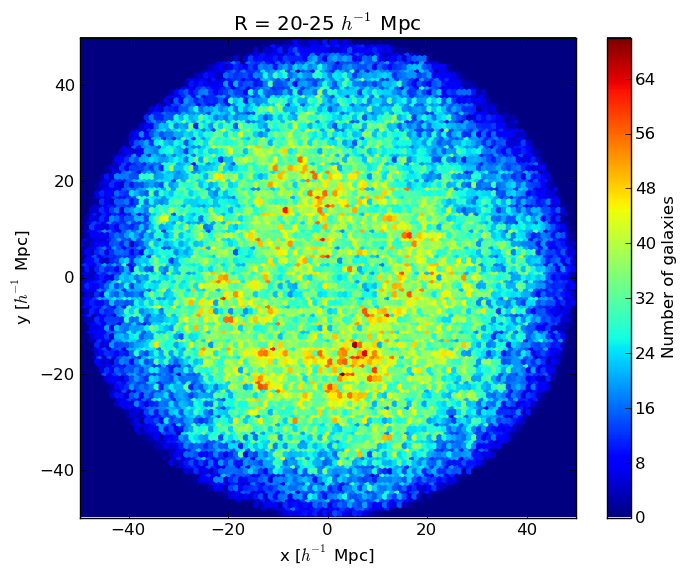}}
  {\includegraphics[type=png,ext=.png,read=.png,width=0.49\textwidth]
    {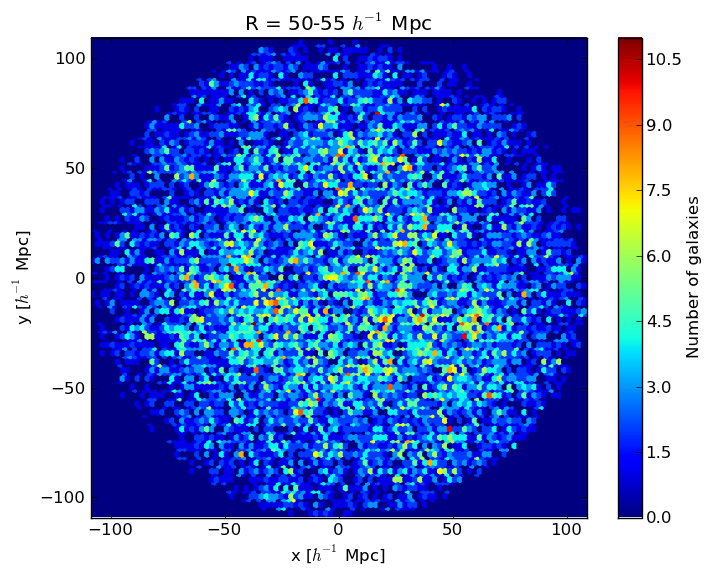}}
  {\includegraphics[type=png,ext=.png,read=.png,width=0.49\textwidth]
    {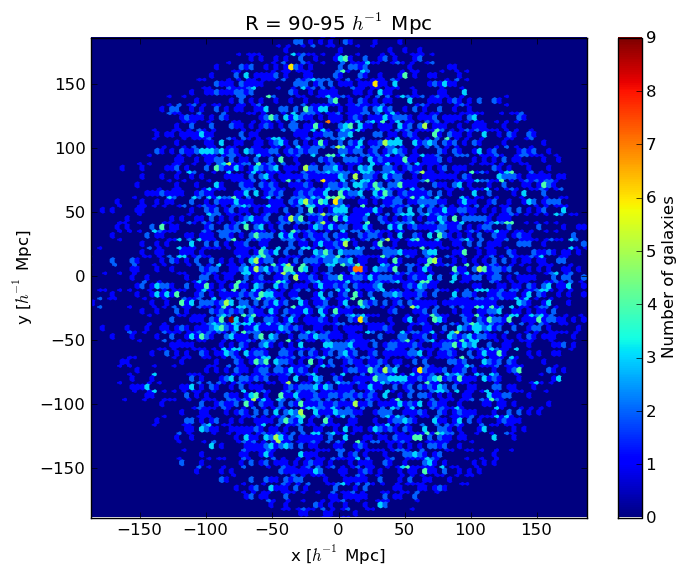}}
  \caption{\emph{Projected distributions of stacked voids.} 
           Here we show two-dimensional 
           projections of stacked voids in various radial size 
           bins. In cases where multiple samples contribute to a 
           stack, we have combined the data.
           }
\label{fig:projections}
\end{figure*}

Our fourth and final
 example derived data product is projections of stacked voids, 
which are useful --- after appropriate rescaling onto the sky --- 
for analyses such as probing the ISW effect 
in CMB observations~\citep[e.g.,][]{Thompson1987, Granett2008} and 
reconstruction of the undistorted void shape.
We show in Figure~\ref{fig:projections} the projected density maps of 
stacked voids for various bin sizes. To construct these we
assume a flat sky approximation within the area of the projected 
stacked void. In cases where multiple samples produce voids within 
the same radial bin, we have combined their voids into a single projection.
We take all galaxies within a spherical region with radius 
twice that of the void effective radius.

As with the radial profiles, we 
see a qualitatively similar distribution that roughly scales with void size:
a minimum-density core, a strongly-defined inner wall, and a 
gradual decrease in density. 
The signal quickly degrades as we move to lower-density samples: 
the voids structure is barely visible in the $50-55$~\hmpc~stack 
and altogether lost to Poisson noise in the $90-95$~\hmpc~stack.
However, these projections reveal structures not apparent in the 
azimuthally-averages radial profiles. Distortions or elongations 
in the wall structure will simply appear as wider compensation regions 
in the profiles but will be immediately noticeable here.
Also, these projections effectively remove any cosmology dependence 
of the void shape, since we have projected them along the 
redshift direction.

%-------------------------------------------------------------------------------
\section{Conclusions}
\label{sec:conclusions}

We have modified the parameter-free 
void finding algorithm {\tt ZOBOV} to account for the survey boundaries 
and internal masks in observational data sets. This prevents 
voids from growing past the survey boundary or into any internal masks.
Thus our approach is more generally applicable to any given 
survey and mask. 
To demonstrate our technique we have
constructed the first public
void catalog using the full extent of the SDSS DR7 spectroscopic 
survey, including the LRGs. 
We combined multiple volume-limited samples of the 
SDSS galaxy catalogs to maximize the number of discovered voids. 
We have produced two catalogs: one catalog that includes all discovered 
voids, including truncated voids near the survey boundaries, and a 
central catalog, which removes voids with questionable shapes and alignments.

The general statistics of our void catalog, such as number 
counts as a function of redshift and 
size distributions, broadly agree with --- but significantly extend --- 
both past analyses of 
observational data~\citep[e.g.,][]{Muller2000,VandeWeygaertR.2011,Pan2011,Patiri2012}
and results from simulations~\citep[e.g.,][]{Dubinski1993, Park2007}.
In addition, radial profiles and projections of stacked 
voids show a qualitatively similar shape across 
the entire sample and agree well with previous efforts.

Due to the relatively poor sampling and the high redshift of the LRG samples,
the topological voids we identify there may not be
strict cosmological features understood as underdensities bounded by 
filaments and walls. 
We may also be overestimating the size of these voids and possibly 
miscalculating their centers. However, the largest voids found in the 
main sample ($\sim 50-60$~\hmpc) overlap with the size distribution of 
voids from the \emph{lrgdim} sample, indicating that there is at least some 
correspondence between the void populations in these samples.
Also, our radial profiles show a qualitatively universal shape in all 
volume-limited subsamples (excepting the \emph{lrgbright} sample), which 
again is a point of evidence that these are truly cosmic voids (note 
especially the similarity in shape in the $50-55$~\hmpc~bin of 
Figure~\ref{fig:profile_samples}). 
In either case, these structures are useful for many kinds of analysis
~\citep[e.g.,][]{Granett2008}.

Our catalogs are useful for many pursuits, including studies of the 
ellipticity distribution of voids, correlations of void positions 
with CMB fluctuations, Alcock-Paczynski tests using the shapes 
of voids in redshift space, and studies of the properties of galaxies 
within voids. We have constructed useful data sets to enable these 
studies, such as catalogs of void galaxies, void stacks of 
various radial sizes, and two-dimensional projections of void 
densities. We have constructed these data sets using both all discovered 
voids and a central void catalog free from survey edge effects. 

We are making our catalogs and data products 
publicly available at~\url{http://www.cosmicvoids.net}. 

%-------------------------------------------------------------------------------
%-------------------------------------------------------------------------------
\section*{Acknowledgments}

PMS and BDW acknowledge support from NSF Grant AST-0908902.
GL acknowledges support from CITA National Fellowship and financial
support from the Government of Canada Post-Doctoral Research Fellowship.
Research at Perimeter Institute is supported by the Government of Canada
through Industry Canada
 and by the Province of Ontario through the Ministry of Research and
Innovation.
DW acknowledges support from NSF Grant AST-1009505.
This material is based upon work supported in part by NSF Grant
 AST-1066293 and the hospitality of the Aspen Center for Physics.

Some of the results in this paper have been derived using the 
\healpix~\citep{Gorski2005} package.

Funding for the Sloan Digital Sky Survey (SDSS) has been provided by the Alfred P. Sloan Foundation, the Participating Institutions, the National Aeronautics and Space Administration, the National Science Foundation, the U.S. Department of Energy, the Japanese Monbukagakusho, and the Max Planck Society. The SDSS Web site is http://www.sdss.org/.

\appendix
\label{sec:layout}
\section{Layout of the Catalog}

We provide the catalog as a single downloadable {\tt gzip}-archived file 
at \url{http://www.cosmicvoids.net}. Most files are in human-readable 
{\tt ASCII} format, with the exception of the raw \zobov-generated 
catalog in binary and the projections which are {\tt NumPy} array files.

There are two top-level directories within the catalog: {\tt figures} and 
{\tt sdss\_dr7}. {\tt sdss\_dr7} contains 
the void catalog and subsequent analysis as 
presented in this paper. The {\tt figures} directory contains data files 
necessary to reproduce the figures found in this paper.

We divide the catalog into directories based on each of our six 
volume-limited subsamples.
Each directory contains the complete \zobov~void catalog (i.e., without any 
size or density cuts). For each sample there are four 
files which describe the voids and their 
member zones and galaxies: one text file which lists the voids, two binary 
files which link voids to zones and zones to particle members, and finally 
a binary file which contains the galaxy positions of the volume-limited sample.
Note that there are two void description files, 
corresponding to the \emph{all} and \emph{central} void catalogs.
To aid in parsing these files, we provide a small exampling catalog 
reading routine in {\tt C}, {\tt dumpVoidParticles}.

Each sample directory also contains two 
subdirectories:~{\tt all} and {\tt central},
which only include voids after size and density cuts.
Each of these subdirectories contains two text files: {\tt centers.txt} and 
{\tt sky\_positions.txt}, which describe the coordinates and properties 
of each void.

The {\tt figures} directory contains data files which can be used to reproduce 
various figures in this paper. The file names indicate the figure and whether
they are from the \emph{central} or \emph{all} catalog.  

An alternate catalog, called {\tt lcdm}, is available and
contains only a raw \zobov~void catalog where galaxy positions are 
converted to real space using a \lcdm~cosmology.

\bibliography{sdssvoids}		
\bibliographystyle{apj}	\nocite{*}

\end{document}